\documentclass[twocolumn, prd]{revtex4}

\usepackage{graphicx}
\usepackage{amsmath}
\usepackage{epstopdf}
\usepackage{times}
\usepackage{tensor}
\usepackage{appendix}
\usepackage{amssymb}
\usepackage[]{natbib}
\usepackage[none]{hyphenat}

\begin{document} 

\title{Stochastic backgrounds in alternative theories of gravity: 
overlap reduction functions for pulsar timing arrays}
\author{Sydney J. Chamberlin}
\affiliation{Center for Gravitation and Cosmology, Department of Physics, 
University of Wisconsin--Milwaukee, P.O. Box 413, Milwaukee, WI 53201, USA} 
\author	{Xavier Siemens}
\affiliation{Center for Gravitation and Cosmology, Department of Physics, 
University of Wisconsin--Milwaukee, P.O. Box 413, Milwaukee, WI 53201, USA}
\begin{abstract}
In the next decade gravitational waves might be detected using a pulsar timing array. 
In an effort to develop optimal detection strategies for  
stochastic backgrounds of gravitational waves in generic metric theories of 
gravity, we investigate the overlap reduction 
functions for these theories and discuss their features. We show that the sensitivity to 
non-transverse gravitational waves is greater than the sensitivity to transverse 
gravitational waves  and discuss the physical origin 
of this effect. We calculate the overlap reduction functions for the 
current NANOGrav Pulsar Timing Array (PTA) and show that the sensitivity to the 
vector and scalar-longitudinal modes can increase dramatically for pulsar pairs with 
small angular separations. For example, the J1853$+$1303--J1857$+$0943 pulsar pair, with 
an angular separation of about $3^\circ$, is about $10^4$ times more sensitive 
to the longitudinal component of the stochastic background, if it is present, 
than the transverse components.
\end{abstract}
\maketitle
\section{Introduction}

General relativity is among the most successful theories 
of physics in the 20th century, passing all current weak-field, 
slow motion tests with flying colors. Progress in 
cosmology and high energy physics over the course of the last 50
years, however, has brought with it questions that may be unanswerable in the context of 
general relativity. The accelerated expansion of the universe, the 
dark matter problem, and inflation 
have led some authors to re-examine general relativity and attempt 
to modify it to explain some of these puzzles. 
Additionally, the incompatibility between general relativity and 
quantum field theory may be an indication that modifications 
to general relativity are necessary. 

A number of alternative theories of gravity have been ~proposed 
to address some of these problems. Those which satisfy the Einstein Equivalence 
Principle are called metric theories of gravity. In these theories, 
the only gravitational fields that may influence matter are the components of the metric 
tensor $g_{\mu \nu}$. Additional fields play the role of generating 
spacetime curvature. Metric theories are grouped broadly into several 
categories:
scalar tensor theories, in which a dynamical scalar field $\phi$ is 
present in addition to the metric (see 
Refs.~\cite{noj-odi, noj-odi-newer, Lobo:2008sg, Alves:2009eg,Capozziello:2007ec,DeFelice:2010aj,%
willbook, brunetti-1999-59, Clifton:2011jh}); 
vector-tensor theories, which contain a dynamic gravitational 
four-vector field in addition to the metric (see Refs.~\cite{willbook,Sagi:2010ei,%
Clifton:2010hz,Clifton:2011jh,Skordis:2009bf}); and bimetric theories, which are 
characterized by ``prior'' geometry contained in dynamical scalar, 
vector or tensor fields (see Refs.~\cite{willbook,Clifton:2011jh,Milgrom:2009gv}). 

Gravitational wave astronomy promises 
not only to open a new observational window on the universe, but also to 
provide a new testing ground for general relativity. In a general metric theory of 
gravity, the six independent components of the Riemann tensor provide up to six 
possible gravitational wave (GW) polarization states, four more than those 
allowed in general 
relativity. Detection of any extra GW polarization states 
would be fatal for general relativity.  A non-detection could be used
put constraints on the parameters of alternative theories of gravity. 

Several international efforts are currently underway to detect GWs. Of these 
the most promising on the 5--10 year timescale are ground-based laser 
interferometers~\cite{Abramovici:1992ah} and pulsar 
timing arrays~\cite{Hobbs:2009yy}, which aim to detect GWs in the $10$--$10^{3}$ Hz 
and $10^{-9}$--$10^{-7}$ Hz ranges, respectively. Potential sources for 
low frequency GWs ($10^{-9}$--$10^{-7}$ Hz) include binary supermassive 
black hole mergers~\cite{Sesana:2008mz}, cosmic superstrings~\cite{Olmez:2010bi}, 
relic gravitational waves from inflation~\cite{Starobinsky:1979ty}, and a first 
order phase transition at the QCD scale~\cite{Caprini:2010xv}. 

Previous work on stochastic backgrounds of gravitational waves in the 
context of alternative theories of gravity has shown that three ground-based 
interferometers are sufficient to disentangle the polarization content 
of a general metric theory of gravity~\cite{nishizawa}. 
For pulsar timing arrays the form of the correlation between 
pulsar pairs as a function of pulsar pair angular separation
depends on the polarization content of the theory~\cite{KJ}. Additionally it has been 
shown that pulsar timing arrays have a greater 
sensitivity to longitudinal and vector polarization 
modes than to transverse modes~\cite{KJ, alvestinto}. 

In this paper we investigate the problem of stochastic GW detection using PTAs in the
context of the optimal statistic.
We compute the expected cross correlations for pulsar timing arrays
for the case of stochastic backgrounds of GWs for any metric theory of 
gravity. The expected cross
correlations are proportional to the so-called overlap reduction function, 
a geometrical factor that captures 
the loss of sensitivity due to detectors not being co-located or aligned.
We explain various features of the overlap reduction functions including the physical 
origin of the increased sensitivity to scalar-longitudinal 
and vector polarization modes. In Section~\ref{sec: background}, we use a coordinate independent 
approach to describe the redshift of pulsar signals from passing GWs. 
In Section~\ref{sec: det-stat} we write the optimal cross-correlation filter by maximizing the signal to noise 
for a pulsar pair, and define the overlap reduction function for GWs of any 
metric theory of gravity. In Section~\ref{sec: FDredshift} we discuss the effect 
of GWs of various polarizations 
on the pulsar-Earth system, and the physical origin of the increased sensitivity to 
longitudinal and shear modes. This effect is 
most easily understood in the frequency domain.
In Section~\ref{sec: ORFstuff}, we write down explicitly the form of the overlap reduction
function for transverse GWs and discuss the form of the function for non-transverse GWs. We 
find that for the scalar-longitudinal and vector (shear) modes, the overlap reduction 
functions are frequency dependent in the ranges of frequencies and distances relevant to 
pulsar timing. This is not the case for the transverse tensor and breathing 
modes. 
In Section~\ref{sec: NGpulsar-ORFs}, we compute the overlap reduction functions for the current 
NANOGrav PTA and show that sensitivity to the scalar-longitudinal and vector (shear) modes 
increases by at least an order of magnitude for nearby pulsar pairs for vector modes, and about 
four orders of magnitude for the  longitudinal mode.  We summarize our results in 
Section~\ref{discussion}. Throughout we work in units where the speed of light $c=1$.

\section{Detecting gravitational waves with a pulsar timing array}\label{sec: background}
\begin{figure}
	\includegraphics[width=3.5in]{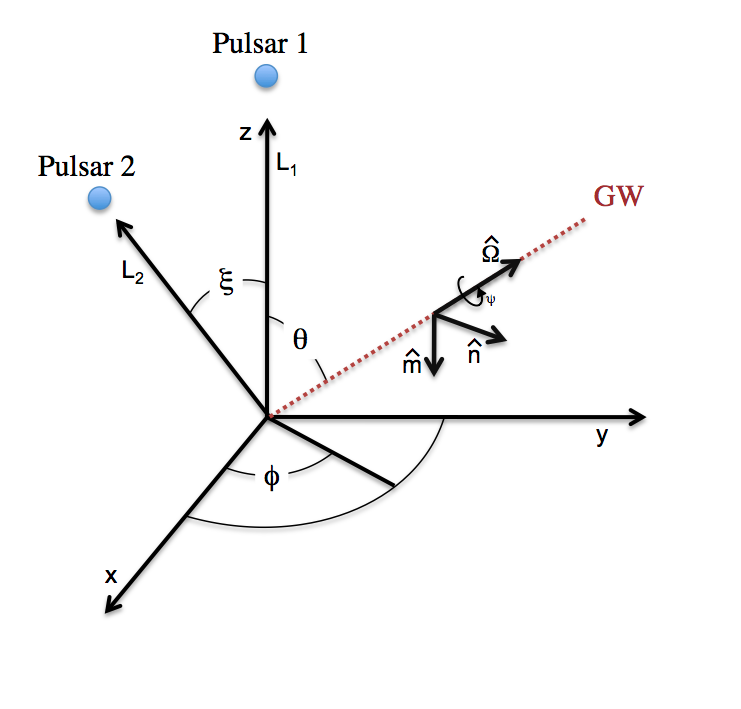}
	\caption{\footnotesize Pulsar positions are given with respect to the Solar System 
barycenter (located at the origin). Here $\theta$ and $\phi$ are the typical 
polar and azimuthal angles (as projected from the position of pulsar 1), and 
pulsar 1 and pulsar 2 are separated by angle $\xi$. A gravitational wave, 
characterized by polarization angle $\psi$, propagates along the 
$\hat{\Omega}$ direction. }\label{fig: pulsardiagram} 
\end{figure}

The radio pulses from pulsars arrive at our radio telescopes at very steady rates.
Pulsar timing experiments exploit this regularity. Fluctuations in the time of arrival of
radio pulses, after all known effects have been accounted for, might be due to the presence
of a GW background. If a GW is present the signal from the pulsar can be red-shifted 
(or blue-shifted). For a GW propagating in the direction $\hat{\Omega}$, 
the redshift of signals from a pulsar in the 
direction $\hat{p}$ is given by~\cite{detweiler,anholm}
\begin{eqnarray}\label{redshift}
	z(t,\hat{\Omega}) = \frac{\hat{p}^i \hat{p}^j}
{2 \left(1+\hat{\Omega} \cdot \hat{p}\right) } 
[h_{ij}(t_p, \hat{\Omega})-h_{ij}(t_e, \hat{\Omega})]
\end{eqnarray}
where $h_{ij}$ is the metric perturbation and $t_p$, $t_e$ represent 
the times the pulse was emitted at the pulsar and the time received at 
the Solar System barycenter, and we have defined
\begin{eqnarray}\label{reddef}
	z(t, \hat{\Omega}) = \frac{\nu_e - \nu_p}{\nu_p}.
\end{eqnarray}
Note that this is the opposite of the sign convention normally used in the 
literature~\cite{detweiler}.
\begin{figure}
	\includegraphics[width=3.5in]{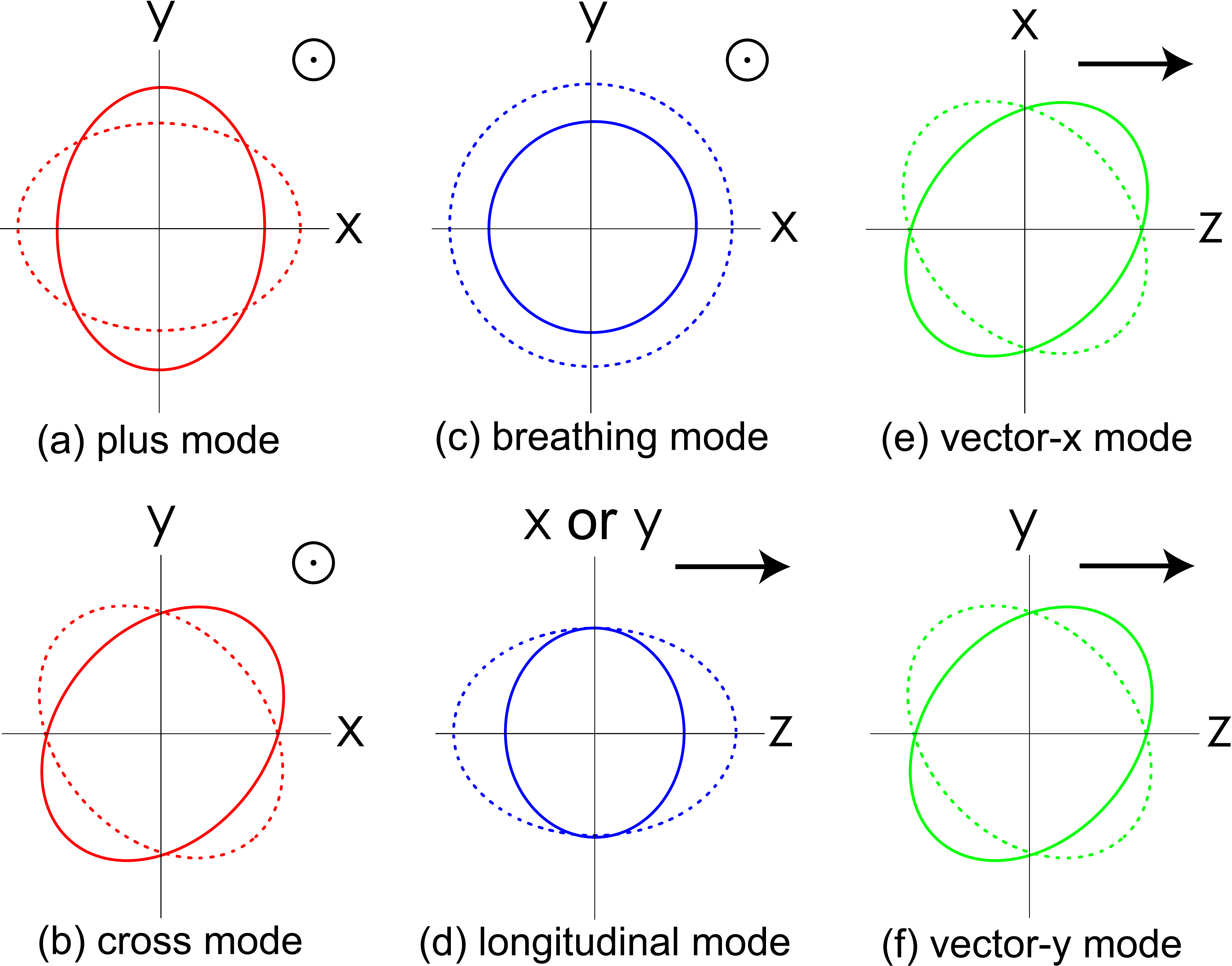}
	\caption{\footnotesize Motion of test masses in response to GWs of the six polarization modes. The plus ($+$), 
	cross ($\times$), and scalar-breathing ($b$) mode GWs are transverse, while the two 
        vector modes ($x$, $y$) 
	and the scalar-longitudinal ($l$) mode GWs are non-transverse. Figure reproduced 
        from Nishizawa et al. \cite{nishizawa} with permission.}\label{fig:gw-pols} 
\end{figure}
Modified gravity theories extend the possible polarization modes of 
GWs present in general relativity -- the plus ($+$) and cross ($\times$) modes-- to 
a maximum of six possible modes. For the two pulsar--Earth system shown in 
Fig.~\ref{fig: pulsardiagram}, the GW coordinate system is given by
\begin{eqnarray}\label{coords}
\hat{\Omega} &=& (\sin{\theta} \cos{\phi},\sin{\theta} \sin{\phi},\cos{\theta}) \nonumber \\
\hat{m} &=& (\sin{\phi}, -\cos{\phi}, 0) \nonumber \\
\hat{n} &=& (\cos{\theta} \cos{\phi},\cos{\theta} \sin{\phi},-\sin{\theta}) \\
\nonumber
\end{eqnarray}
where, relative to \cite{nishizawa}, we have fixed the GW polarization angle $\psi = -\pi/2$ 
to agree with the conventions in~\cite{allenromano}. From \eqref{coords}, 
the GW polarization tensors can be constructed \cite{PRL30, KJ, nishizawa, tintoalves2010, alvestinto}
\begin{eqnarray}\label{poltensors}
	 \epsilon_{ij}^{+} = \hat{m} \otimes \hat{m} - \hat{n} \otimes \hat{n}, 
&  \epsilon_{ij}^{\times} = \hat{m} \otimes \hat{n} + \hat{n} \otimes \hat{m} \nonumber \\
	 \epsilon_{ij}^{b} = \hat{m} \otimes \hat{m} + \hat{n} \otimes \hat{n}, 
& \epsilon_{ij}^{l} = \hat{\Omega} \otimes \hat{\Omega} \nonumber \\
	 \epsilon_{ij}^{x} = \hat{m} \otimes \hat{\Omega} + \hat{\Omega} \otimes \hat{m}, 
& \epsilon_{ij}^{y} = \hat{n} \otimes \hat{\Omega}+\hat{\Omega} \otimes \hat{n} \nonumber \\
\end{eqnarray}
where $\otimes$ is the tensor product and $\hat{\Omega}$ is the direction of 
GW propagation. Here, $x$ and $y$ correspond to the vector (spin-1) polarization 
modes while $b$ and $l$ correspond to the scalar (spin-0) breathing and longitudinal 
modes, respectively. The plus, cross and breathing modes are characterized by 
transverse GW propagation, while the longitudinal and vector (or shear) 
modes are non-transverse in nature (see Fig.~\ref{fig:gw-pols}).

Defining the antenna patterns as
\begin{eqnarray}\label{antpatts}
	F^A(\hat{\Omega})  = \epsilon_{ij}^A(\hat{\Omega}) 
\frac{\hat{p}^i \hat{p}^j}{2 (1+\hat{\Omega} \cdot \hat{p})},
\end{eqnarray}
the Fourier transform of \eqref{redshift} may be written as \cite{KJ, anholm, tintoalves2010}
\begin{eqnarray}\label{FTantpatts}
	\tilde{z}(f, \hat{\Omega}) = \left(e^{-2 \pi i f L 
(1+\hat{\Omega} \cdot \hat{p})} -1 \right) \sum_A \tilde h_A(f,\hat{\Omega})F^A(\hat{\Omega})
\end{eqnarray}
where the sum is over all possible GW polarizations: 
$A~=~+, 
\times, x, y, b,l$, and $L$ is the distance to the pulsar. 

\begin{figure*}[!]
	\begin{minipage}[b]{0.4\textwidth}
	\centering
		\includegraphics[width=\textwidth]{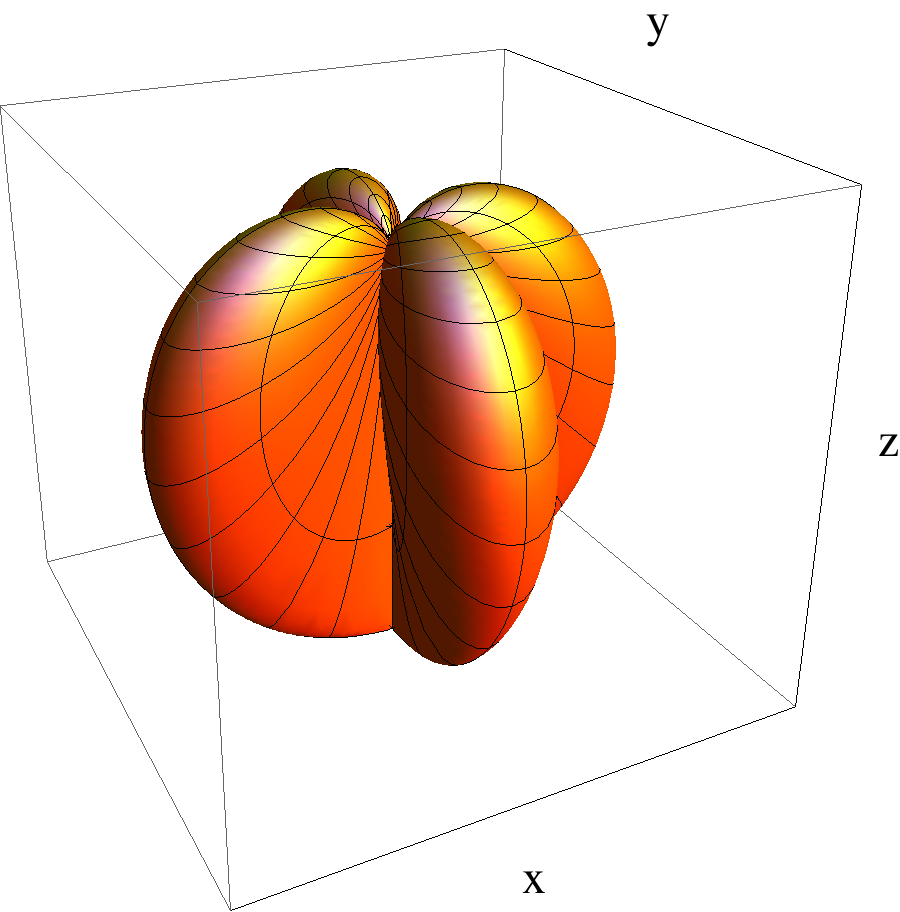}
		\centerline{(a)}
	\end{minipage}
	~
	~
	\begin{minipage}[b]{0.4\textwidth}
	\centering
		\includegraphics[width=\textwidth]{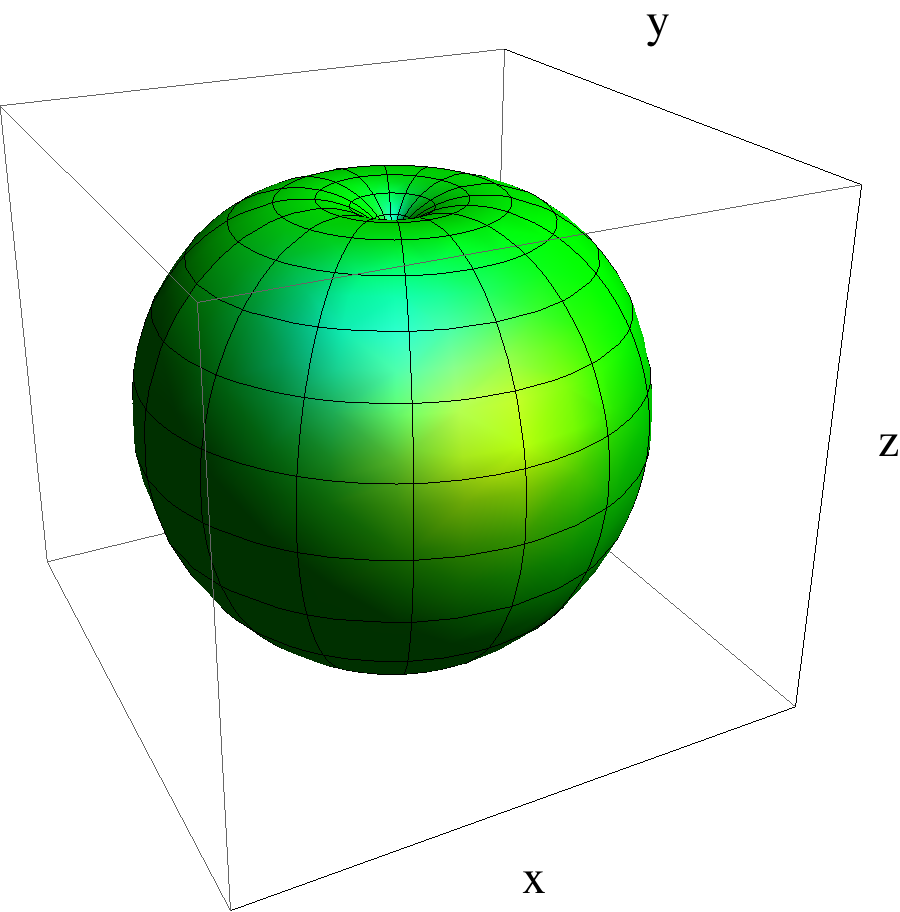}
		\centerline{(b)}
	\end{minipage}
	
	\
	
	\begin{minipage}[b]{0.4\textwidth}
	\centering
		\includegraphics[width=\textwidth]{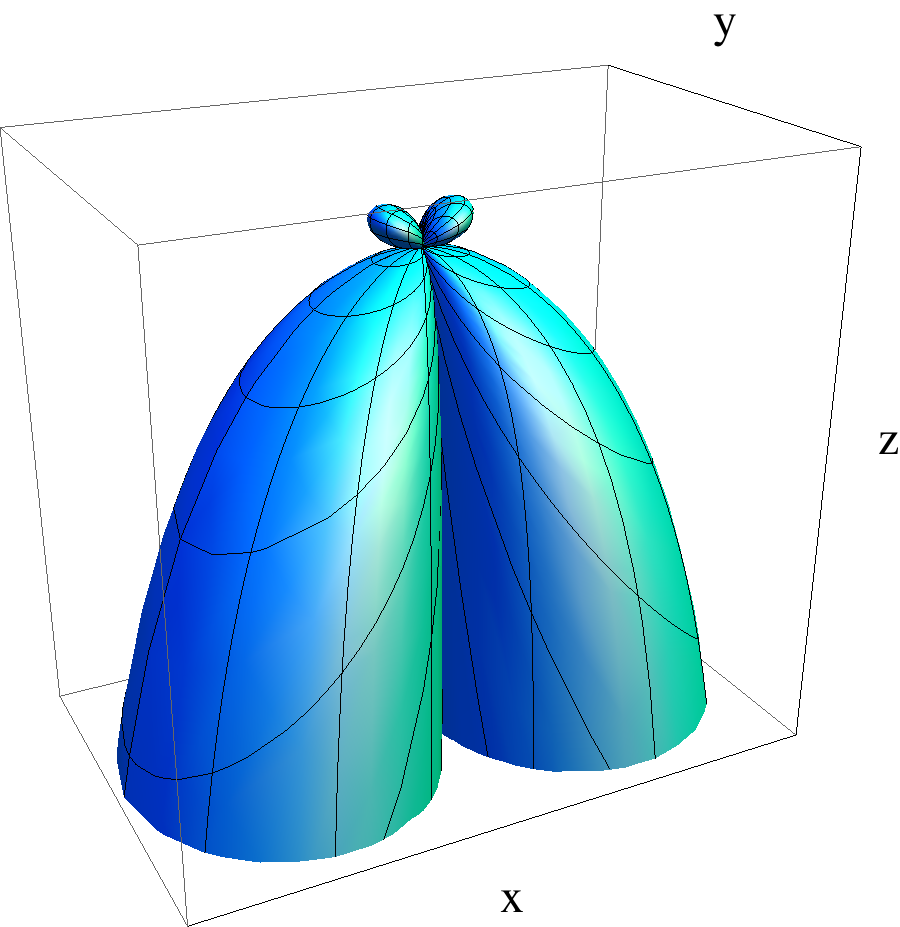}
		\centerline{(c)}
	\end{minipage}
	~
	~
	\begin{minipage}[b]{0.4\textwidth}
	\centering
		\includegraphics[width=\textwidth]{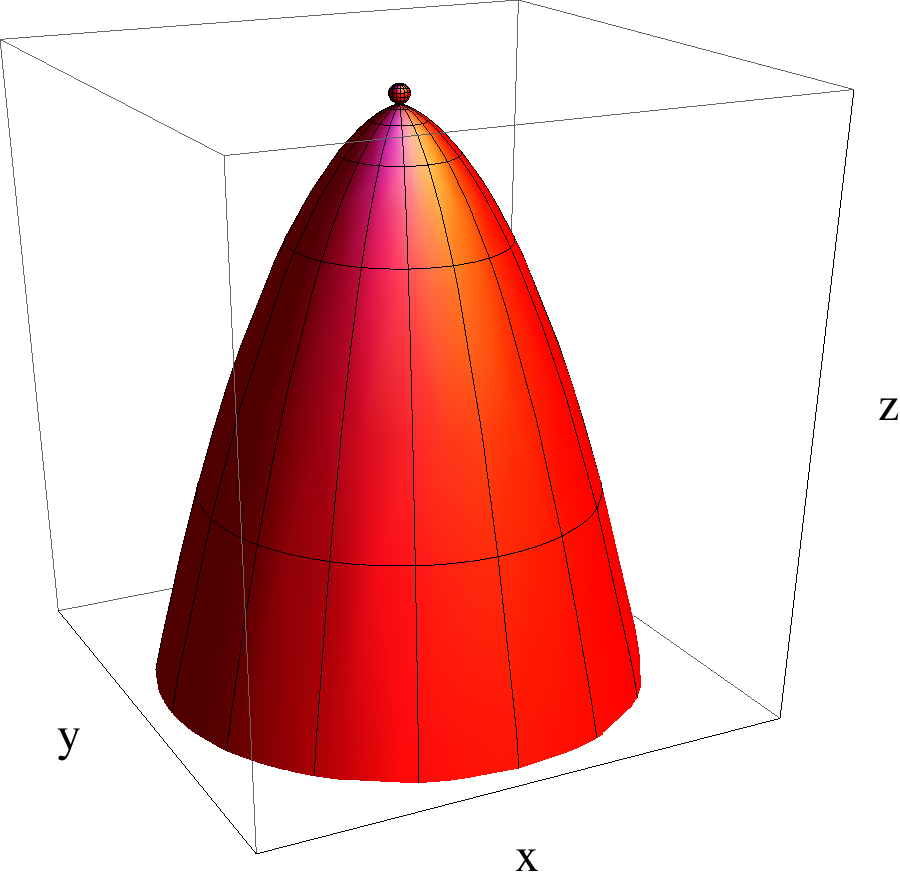}
		\centerline{(d)}
	\end{minipage}
	\caption{\footnotesize Antenna patterns \eqref{antpatts} for plus/cross (a), breathing (b), vector-x/vector-y (c), and longitudinal 
	(d) polarization modes. Note that the cross and vector-y modes are
	identical to plus and vector-x, respectively, but rotated by 45 degrees and thus do not appear separately here. 
	In this figure, the GW propagates in the positive z-direction with the Earth at the origin, 
	and the antenna pattern depends on the pulsar's direction, specified 
	by polar angle $\theta_p$ and azimuthal angle $\phi_p$. Exact 
	expressions corresponding to each figure may be found in Appendix~\ref{sec: appA}: 
	\eqref{plusz} for the plus mode, \eqref{breathez} for the breathing mode, \eqref{xz} for the vector-x mode, and \eqref{longz} for the longitudinal mode.
	Note that fixing the GW propagation direction while allowing the pulsar position to change 
	is analogous to fixing the pulsar position while 
	allowing the direction of GW propagation to change 
       (there is an inherent degeneracy in the GW polarization 
	angle and the pulsar's azimuthal angle $\phi_p$). }\label{fig: antpatt} 
\end{figure*}

%
The actual quantity measured in pulsar timing experiments is the 
timing residual, which is defined as the difference between the 
actual and expected time of arrival (TOA) of a pulse: 
\begin{eqnarray}
	R(t) = \rm{TOA_{\hspace{1mm }actual}}-\rm{TOA_{\hspace{1mm } expected}}.
\end{eqnarray}
The expected TOA for a pulse is modeled and includes daily and yearly motion of 
the Earth, the position and proper motion of the pulsar, motion about a 
binary companion (if applicable), etc. 
The timing residual can be obtained by integrating the redshift in 
time~\cite{detweiler}.

In Fig.~\ref{fig: antpatt} we plot the antenna patterns for the various GW 
polarization modes in a system where the GW's direction of propagation  
is fixed and the pulsar's position is varied (see Appendix~\ref{sec: appA}, 
Eqns. \eqref{plusz}, \eqref{breathez}, \eqref{xz} and \eqref{longz} for details), 
as is usually done in the literature.

\section{GW detection statistic}\label{sec: det-stat}

In this section we introduce the optimal cross correlation statistic~\cite{allenromano,anholm} for
stochastic background searches. The optimal cross-correlation statistic involves 
the calculation of the overlap reduction 
function, a geometrical factor that 
characterizes the loss of sensitivity due to detectors not being co-located
or aligned. We will show how the overlap reduction function is computed for non-transverse modes.
We follow the analysis (for General Relativity) of Allen and Romano~\cite{allenromano}.

The plane wave expansion for a GW perturbation 
propagating in the direction $\hat{\Omega}$ is given by \cite{allenromano}
\begin{eqnarray}
h_{ij}(t, \vec{x})&=&\sum_A \int_{-\infty}^{\infty} df \int_{S^2} d\Omega e^{2 \pi if (t- \hat{\Omega} \cdot \vec{x})} h_A(f, \hat{\Omega}) \epsilon^A_{ij} (\hat{\Omega}) \nonumber \\
\,
\end{eqnarray}
where $i, j$ are spatial indices, the sum is over all six polarization states, and the 
Fourier amplitudes $h_A (f, \hat{\Omega})$ are complex functions satisfying 
$h_A(-f, \hat{\Omega}) = h_A^*({f, \hat{\Omega}})$. 
A stochastic background of GWs is produced by a large number of weak, independent, unresolvable sources. The energy density of this background per unit logarithmic frequency is given by 
\begin{eqnarray}
	\Omega_{\rm{gw}}(f) = \frac{1}{\rho_{\rm{critical}}} \frac{d \rho_{\rm{gw}}}{d \ln{f}}
\end{eqnarray}
where $d \rho_{\rm{gw}}$ is the energy density of the gravitational waves and $\rho_{\rm{critical}}$ is the critical energy density required to close the universe,
\begin{eqnarray}
	\rho_{\rm{critical}} = \frac{3 H_0^2}{ 8 \pi G} 
\end{eqnarray}
where $H_0$ is the Hubble constant.

The characteristic strain spectrum, $h_c(f)$, is typically given a power-law dependence on frequency so that 
\begin{eqnarray}\label{hc}
	h_c(f) &=& A \left( \frac{f}{\rm{yr}^{-1}} \right)^{\alpha}.
\end{eqnarray}
It may also be expressed in terms of the energy density of the background per unit logarithmic frequency, $\Omega_{\rm{gw}}(|f|)$:
\begin{eqnarray}
	h^2_c(f) &=& \frac{3 H_0^2}{2 \pi ^2} \frac{1}{f^2} \Omega_{\rm{gw}}(|f|).
\end{eqnarray}

For an isotropic stochastic background of GWs, the signal appears in the data as correlated noise between measurements from different pulsars. The $i^{th}$ data set is of the form 
\begin{eqnarray}
	s_i(t) &=& z_i(t) + n_i(t)
\end{eqnarray}
where $z_i(t)$ corresponds to the unknown GW signal and $n_i(t)$ to noise 
(assumed in this case to be stationary and Gaussian). Because the 
signal is assumed to be much smaller than the noise, the properties of the 
noise determine the variance. We can express these properties in the 
frequency domain as
\begin{eqnarray}
	\langle \tilde{n}_i(f) \rangle &=& 0 \\
	\langle \tilde{n}^*_i(f) \tilde{n_j}(f') \rangle &=& \frac{1}{2} \delta(f-f')P_i(|f|)
\end{eqnarray}
where we have introduced the one-sided noise power spectrum $P_i(|f|)$. 

The cross-correlation statistic is defined as 
\begin{eqnarray}\label{ccstatQ}
	S = \int_{-T/2}^{T/2} dt \int_{-T/2}^{T/2} dt' s_i(t) s_j(t') Q(t-t')
\end{eqnarray}
where $Q(t-t')$ is the filter function. 
 The optimal filter is determined by maximizing the expected signal-to-noise ratio
\begin{eqnarray}\label{snr}
	\rm{SNR} = \frac{\mu}{\sigma}.
\end{eqnarray}
Here $\mu$ is the mean $\langle S \rangle$ and $\sigma$ is the square root of the variance $\sqrt{ \langle S^2 \rangle - \langle S \rangle^2}$. 

In the frequency domain \eqref{ccstatQ} becomes
\begin{eqnarray}
	S = \int_{-\infty}^{\infty} df \int_{-\infty}^{\infty} df' \delta_T(f-f') \tilde{s}^*_i(f) \tilde{s}_j(f') \tilde{Q}(f'),
\end{eqnarray}
and the mean $\mu$ is
\begin{eqnarray}
	\mu &=& \int_{-\infty}^\infty df \int_{-\infty}^\infty df' \, \delta_T(f-f') \langle \tilde{z}_i^{*}(f) \tilde{z}_j(f') \rangle \tilde{Q}(f') \nonumber \\
\end{eqnarray}
where $\delta_T$ is the finite time approximation to the delta function 
$$
	\delta_T(f)=\frac{\sin{\pi f t}}{\pi f}.
$$
The assumption that the background is unpolarized, isotropic, and stationary implies that the expectation value of the Fourier 
amplitudes $h_A(f, \hat{\Omega})$ must satisfy \cite{allenromano, anholm}
\begin{eqnarray}\label{FA-exp}
	\langle h^*_A (f, \hat{\Omega}) h_{A'} (f',\hat{\Omega}')\rangle &=& \frac{3 H_0^2}{32 \pi^3} \delta^2(\hat{\Omega}, \hat{\Omega}') \delta_{AA'}\\ 
	&\times& \delta(f-f')  |f|^{-3} \Omega_{gw}(|f|) \nonumber
\end{eqnarray}
where $\delta^2(\hat{\Omega}, \hat{\Omega}')$ is the covariant Dirac delta function on the two-sphere. 
With the demand \eqref{FA-exp} in place, the expectation value of the signals $z_i(f)$ may be written as
\begin{eqnarray}
	\langle \tilde{z}_i^*(f) \tilde{z}_j (f') \rangle &=& \frac{3 H_0^2}{32 \pi^3 } \frac{1}{\beta} \delta(f-f') |f|^{-3} \\
	&\times& \Omega_{\rm{gw}}(|f|) \Gamma(|f|). \nonumber
\end{eqnarray}
Here $\beta$ is a normalization factor and we define~\cite{anholm}
\begin{eqnarray}\label{orfs2}
	\Gamma(|f|) &=& \beta \sum_A \int_{S^2} d\Omega \, (e^{2 \pi i f L_i (1+\hat{\Omega} \cdot \hat{p}_i)}-1)\\
	&\times& (e^{-2 \pi if L_j (1+\hat{\Omega} \cdot \hat{p}_j)}-1)  F_i^A(\hat{\Omega}) F_j^A(\hat{\Omega}) \nonumber
\end{eqnarray}
where the sum is over all possible GW polarizations, and the exponential phase terms 
correspond to the pulsar term in the time domain. 

The optimal filter is given by~\cite{allenromano,anholm}
\begin{equation}
\tilde Q(f) \propto \frac{\Omega_{\rm{gw}}(f)\Gamma(f)}{|f|^3 P_i(f)P_j(f)},
\label{optimalfilter}
\end{equation}
where $P_i(f)$ and $P_j(f)$ are the power spectra for the $i$th and $j$th 
pulsar redshift time series that are being cross-correlated (see Eq.~\ref{ccstatQ}).

In general relativity, for the
frequency and distance ranges appropriate to pulsar timing experiments (i.e. for $f \gg 1/L$), 
the overlap reduction function $\Gamma(f)$ approaches a constant which is 
only a function of the angular separation between the two pulsars. This constant is proportional 
to the value of the Hellings-Downs curve for the angle between the pulsars~\cite{hellingsdowns, anholm}. 
We will see that for longitudinal modes and for tensor modes the overlap reduction 
function remains frequency dependent, even for $f \gg 1/L$, and is considerably larger than for the transverse 
modes. This indicates an increased sensitivity to such modes. 
To understand the physical origin of the increased sensitivity 
we first discuss the effect of GWs in the more simple case of a single 
pulsar-Earth baseline.

\section{GW induced redshift on the pulsar-Earth system}\label{sec: FDredshift}
In this section we will study the redshifts induced by GWs of different 
polarizations on the pulsar-Earth system. 
From \eqref{FTantpatts}, the redshift induced by this GW may be written as 
\begin{eqnarray}\label{gen-redshift}
	\tilde{z}_A(f,\hat{\Omega}) = \left(e^{- 2 \pi i f L 
(1 + \hat{\Omega} \cdot \hat{p})} -1 \right)
\frac{p^i p^j}{2 (1 + \hat{\Omega} \cdot \hat{p})} \epsilon_{ij}^A(\hat{\Omega}) \tilde h_A.
\nonumber
\\
\end{eqnarray}
The factor of $1/ 2 (1 + \hat{\Omega} \cdot \hat{p})$ comes 
from the relationship
between the affine parameter $\lambda$ and time $t$ (see Eq.~\eqref{affineparam}), and 
$\tilde h_A=\tilde h_A(f, \hat \Omega)$. 

In the region where the GW direction, $\hat \Omega$ and the pulsar direction, $\hat p$ 
are anti-parallel, \eqref{gen-redshift}
appears to become singular due to the $1+\hat{\Omega}\cdot\hat{p}$ term in the denominator 
(note that the derivative of $h_A$ with respect to the affine parameter vanishes 
in this limit; see ~\eqref{affineparam}). 
There is in fact no divergence in the redshift induced. In this regime the exponential 
can be Taylor expanded and 
the  $1+\hat{\Omega}\cdot\hat{p}$ term in the denominator cancels.

A Taylor expansion of \eqref{gen-redshift} can be performed in two cases. 
In the first, when $fL\ll 1$, the metric perturbation is
the same at the pulsar and at the Earth. This case is often referred to as the 
long wavelength limit. In the second, 
when 
$$1+\hat{\Omega}\cdot\hat{p} \ll \frac{1}{f L},$$ 
the pulse's
direction of propagation and the GW are nearly parallel (i.e. the GW is 
coming from a direction near the pulsar). In this case the 
metric perturbation at the pulsar when the pulse is 
emitted, and on Earth when the pulse is received, are also nearly 
the same. This is often described in the literature 
in terms of the pulse ``surfing'' the gravitational wave. 

The surfing description, combined with Eq.~\eqref{redshift}, might lead one 
to incorrectly conclude that the effect of the GW should cancel in this case 
because the metric perturbations at the Earth and the pulsar are the same, 
despite the divergent $1/(1+\hat{\Omega}\cdot\hat{p})$ term in the redshift. 
In fact, a delicate cancellation occurs with the divergent term in the 
denominator which is only manifest in the frequency domain.
Let the pulse direction and the gravitational wave direction be nearly parallel so that 
$\hat{\Omega}\cdot\hat{p} = -1 + \delta$, where $\delta \ll 1$. Then as in 
\cite{anholm, tintoalves2010} we obtain
\begin{eqnarray}\label{expanded-zA}
	 \tilde{z}_A(f,\hat{\Omega}) \sim - \pi i f L p^i p^j \epsilon_{ij}^A \tilde{h}_A.
\end{eqnarray}
The redshift is proportional to $fL$, but for finite $\delta$ increases only to the point where
the argument of the exponential in \eqref{gen-redshift} can no longer be Taylor expanded, at which 
point it becomes an oscillatory function of $fL$.  
Whether the redshift is finite in the  
$\delta \rightarrow 0$ limit depends on the projection term $p^i p^j \epsilon_{ij}^A h_A$.
As we will see, the vanishing contribution for the tensor modes of general relativity occurs solely 
because of the transverse nature of these waves, and is unrelated to the ``surfing'' effect. 
For longitudinal modes the projection term does not vanish, and the increase in 
sensitivity to such modes originates from GWs that come from directions 
near the pulsar. To better understand this, we will look at the behavior of the redshifts 
induced by GWs of various modes.

The redshift for a longitudinal mode GW perturbation is 
\begin{eqnarray}\label{lmred}
	\tilde{z}_l(f,\hat{\Omega}) &=& \frac{\cos^2{\theta}}{2 (1+\cos{\theta})}
(e^{-2 \pi i f L (1+\cos{\theta})}-1)  \tilde{h}_l,
\end{eqnarray}
while the redshift for a plus mode GW perturbation is
\begin{eqnarray}\label{pmred}
	\tilde{z}_+(f,\hat{\Omega}) &=& \frac{-\sin^2{\theta}}{2 (1+\cos{\theta})} 
(e^{-2 \pi i f L (1+\cos{\theta})}-1) \tilde{h}_+.
\end{eqnarray} 
Here we note that the geometrical factor in the redshift for the transverse breathing mode differs 
from \eqref{pmred} only by a sign, and our analysis of \eqref{pmred} applies equally to the 
breathing mode. 

\begin{figure*}
	\begin{minipage}[b]{0.45\textwidth}
	\centering
	\includegraphics[width=\textwidth]{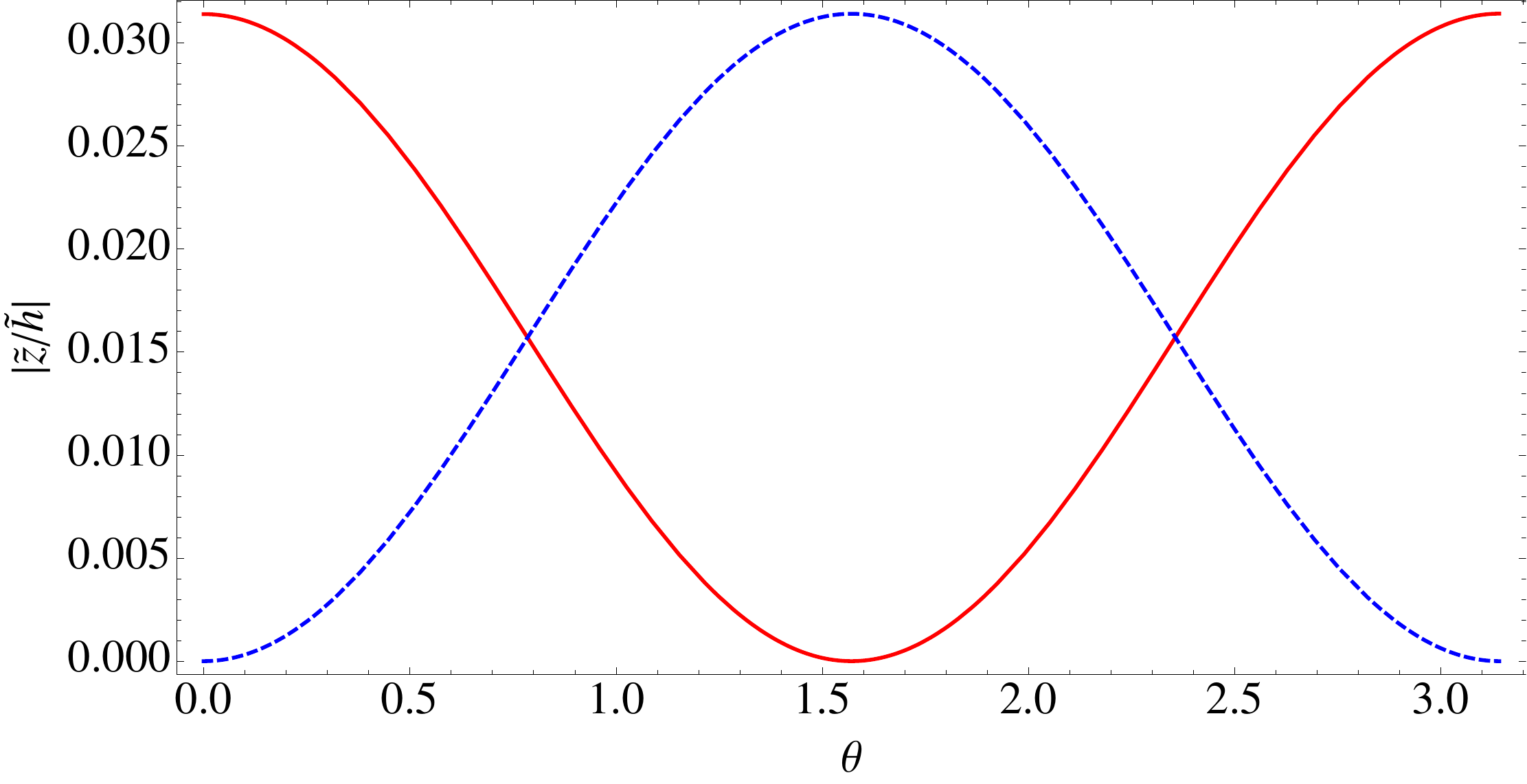}
	\centerline{(a)}
	\end{minipage}
	~ 
	~
	\begin{minipage}[b]{0.45\textwidth}
	\centering
	\includegraphics[width=\textwidth]{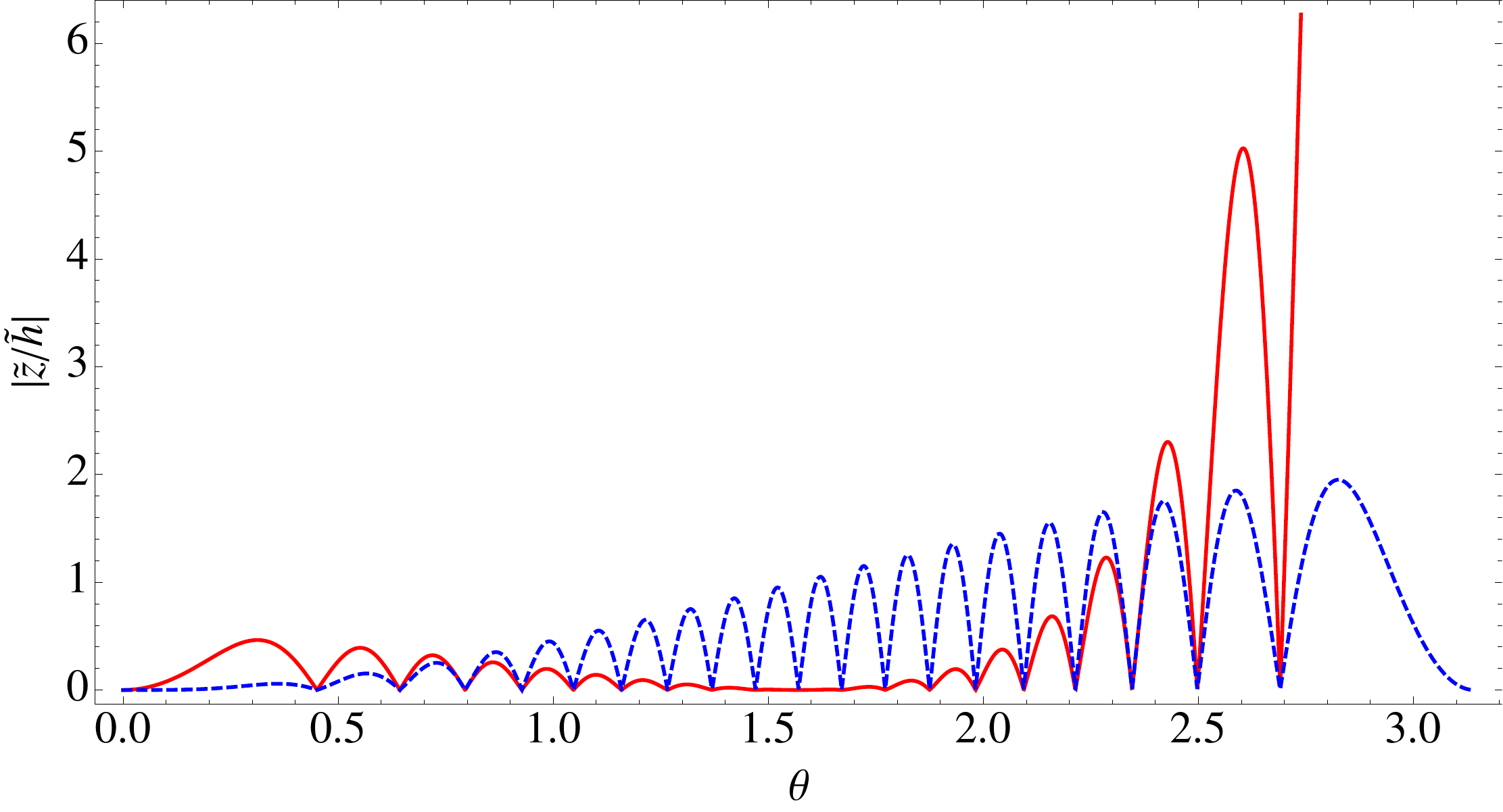}
	\centerline{(b)}
	\end{minipage}
	\caption{\footnotesize(color online) Plots of $|\tilde{z}(f,\hat{\Omega})/\tilde{h}|$ 
for the $+$-mode (dashed blue) and the longitudinal mode (solid red). We show these 
for $fL = 10^{-2}$ (a), a value of $fL$ in the long wavelength limit, and (b) $fL=10$, a value 
of $fL$ typical of pulsar timing experiments. In the 
regime of pulsar timing experiments the sensitivity is largest for GW directions near the 
pulsar $\theta \approx \pi$ for both polarizations. In the longth wavelength limit, $fL \ll 1$, the 
pulsar-Earth system is most sensitive to $+$-mode GWs coming from the equator, and longitudinal GWs
from the poles.} 
\label{fig:plusandlong} 
\end{figure*}

%
%
In Fig.~\ref{fig:plusandlong} we plot the geometrical and phase factor $|\tilde{z}(f,\hat{\Omega})/\tilde{h}|$ 
for both the $+$-mode and the longitudinal mode. We plot these for a value of $fL$ in the long wavelength 
limit ($fL = 10^{-2}$), and for a value in the regime of pulsar timing experiments ($fL=10$). In the 
regime of pulsar timing experiments the sensitivity is largest for GW directions near the 
pulsar $\theta \approx \pi$ for {\it both} polarizations. Although we do not show it here 
the same is true for all other polarization modes. In the long wavelength limit, $fL \ll 1$, the 
pulsar-Earth system is most sensitive to $+$-mode GWs coming from the equator, and longitudinal GWs
from the poles.

As discussed above, these redshifts appear to become singular when $\theta \rightarrow \pi$, but 
the pulsar term may be Taylor expanded.
Let $\theta = \pi - \delta$, where $\delta \ll 1$. Then 
\begin{eqnarray}\label{expanded-zl}
	 \tilde{z}_l(f,\hat \Omega) \sim \pi i f L (1-\delta^2) \tilde{h}_l
\end{eqnarray}
for the longitudinal case, while
\begin{eqnarray}\label{expanded-zp}
	 \tilde{z}_+(f,\hat \Omega) \sim \pi i f L \delta^2 \tilde{h}_+
\end{eqnarray}
for the plus mode. In the limit as $\delta \rightarrow 0$, $\tilde{z}_+$ 
vanishes while $\tilde{z}_l$ becomes proportional to $fL$. The vanishing redshift
of $\tilde{z}_+$  is therefore due to the transverse nature of the mode, and does 
not occur for $\tilde{z}_l$, even though in both cases 
the pulse is ``surfing'' the GW.
In the time domain, in the $\theta \approx \pi$ region, the redshift for both modes goes as
\begin{eqnarray}\label{velterm}
	z_{l,+}(t, \hat{\Omega}) \propto L \dot{h}_{l, +}.
\end{eqnarray}
One may readily identify the right hand side of \eqref{velterm} as a velocity.
The interpretation of this result is that, in this limit, the redshift is 
proportional to the relative velocity of the pulsar-Earth system. 
The velocity of the pulsar when the pulse is 
emitted in this limit is approximately 
equal and opposite to the velocity of the Earth when the pulse is 
received. 

An identical analysis for the shear GW modes produces analogous results. 
Starting from \eqref{FTantpatts}, the redshift 
for the vector-y mode goes as 
\begin{eqnarray}
	\tilde{z}_y(f,\hat \Omega) &=& -\frac{\cos{\theta} 
\sin{\theta}}{(1+\cos{\theta})} (e^{-2 \pi i f L (1+\cos{\theta})}-1)  h_y.
\end{eqnarray}
The small $\delta$ expansion yields
\begin{eqnarray}
	\tilde{z}_y(f,\hat \Omega) \sim  - 2 \pi i f L \delta 
\left(1-\frac{\delta^2}{2}\right) h_y.
\end{eqnarray}

Relative to the longitudinal mode the redshift of vector modes is smaller by 
a factor of $\delta$ and vanishes as $\delta \rightarrow 0$, but it is still larger
than the transverse modes by a factor of $1/\delta$.

The same behavior is not present in other sky locations. If the GW 
propagates perpendicular to the pulsar-Earth line 
($\theta~=~\pi/2~+~\delta$), then up to second order in $\delta$ 
the redshifts
\begin{eqnarray}
	\tilde{z}_l &=&\frac{\delta^2}{2 (1-\delta)}  
\left( e^{- 2\pi i f L (1-\delta)} -1\right) \mbox{  (longitudinal)} \\
	\tilde{z}_+ &=&\frac{- \left( 1 - \delta^2 \right) }{2(1-\delta)} 
\left( e^{- 2\pi i f L (1-\delta)} -1\right)  \mbox{  (plus)}\\
	\tilde{z}_y &=&\frac{\delta \left(1-\delta^2/2 \right)}{(1-\delta)} 
\left( e^{- 2\pi i f L (1-\delta)} -1\right)  \mbox{  (shear)} 
\end{eqnarray}
are obtained.  In this case for small $\delta$ the exponential cannot be 
expanded unless $f L \ll 1$. For this sky location the redshift is always an oscillatory function of $fL$.
The pulse comes across different phases of the GW as it propagates toward Earth.

To summarize, one can see that the surfing effect does not lead to a 
vanishing response of the pulsar-Earth system to GW waves 
coming from $\theta = \pi$. For the tensor and scalar-breathing modes,
it is the transverse nature of GWs that is responsible for the vanishing 
response. For the scalar-longitudinal modes the response does not vanish---in fact, 
the response increases with both frequency and pulsar distance. 
For the vector modes the response does vanish, but more slowly than for the 
transverse modes. For {\it all} GW modes from directions near  $\theta = \pi$,
the redshift increases monotonically up to some limiting frequency 
beyond which the Taylor series expansion of the pulsar term 
which leads to Eqs. \eqref{expanded-zl} and  \eqref{expanded-zp} 
can no longer be performed. 

We now discuss the implications of 
this effect on the overlap reduction functions. 

\section{Overlap reduction functions}\label{sec: ORFstuff}
As discussed in Section~\ref{sec: det-stat}, the overlap reduction 
function for the two pulsars in Fig.~\ref{fig: pulsardiagram} is 
equal to 
\begin{eqnarray}\label{orfs}
	\Gamma(|f|) &=& \frac{3}{4 \pi} \sum_A \int_{S^2} d\Omega 
(e^{2 \pi i f L_1 (1+\hat{\Omega} \cdot \hat{p}_1)}-1)\\*
	&\times& (e^{-2 \pi if L_2 (1+\hat{\Omega} \cdot \hat{p}_2)}-1)  
F_1^A(\hat{\Omega}) F_2^A(\hat{\Omega}) \nonumber \\
	&=& \Gamma_{+}(|f|) + \Gamma_{\times}(|f|)+\Gamma_{b}(|f|)+\Gamma_{l}(|f|)\\*
	&+& \Gamma_{x}(|f|)+\Gamma_{y} \nonumber (|f|)
\end{eqnarray}
where all possible GW polarizations are allowed. It is advantageous 
to consider each term in the sum~\eqref{orfs} separately since 
various gravity theories may have different polarization content~\cite{willbook, nishizawa,Lobo:2008sg, Alves:2009eg,Capozziello:2007ec,DeFelice:2010aj, brunetti-1999-59, Clifton:2011jh,Sagi:2010ei,Clifton:2010hz,Skordis:2009bf,Milgrom:2009gv}. 
\begin{figure}
	\includegraphics[width=3in]{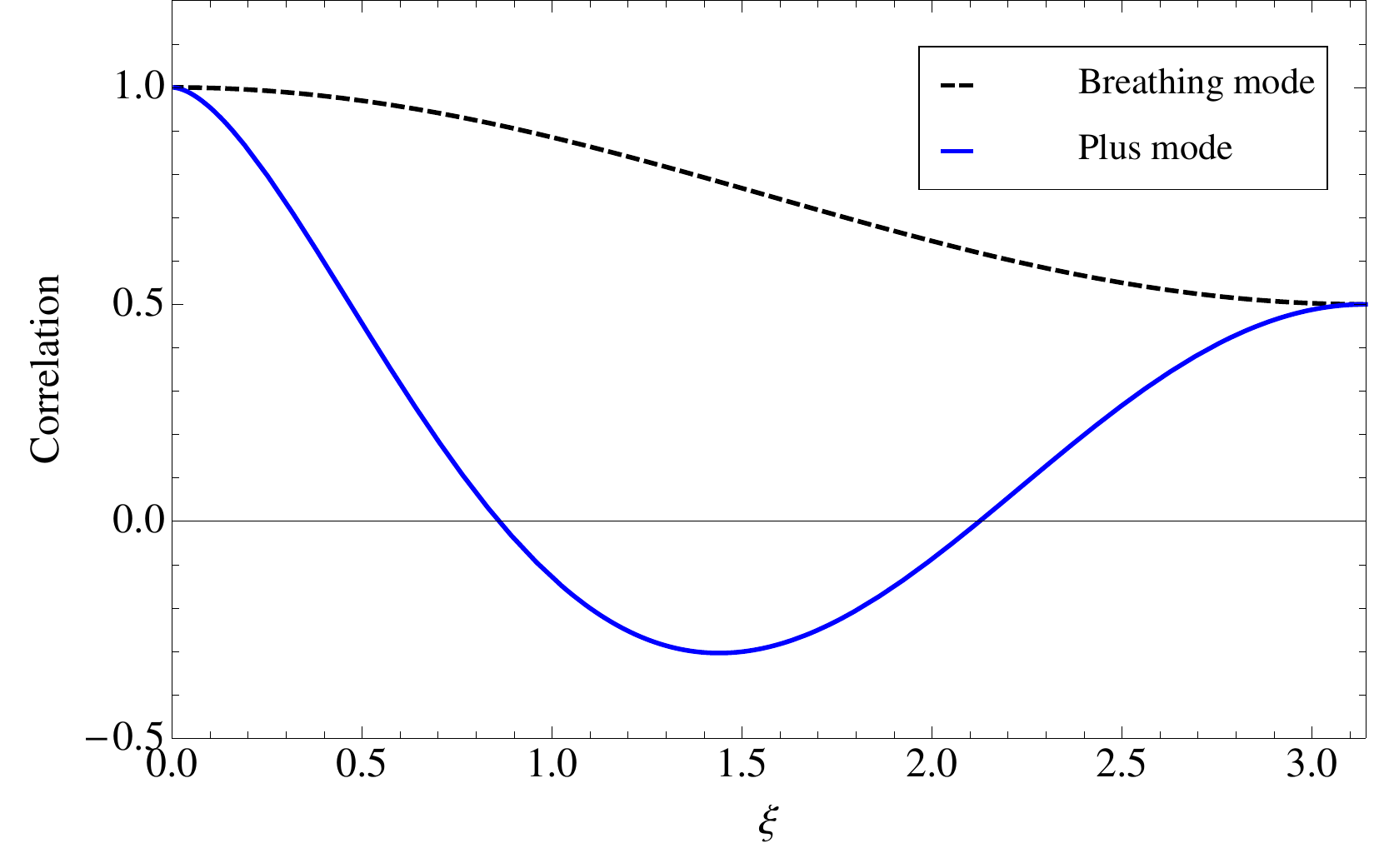}
	\caption{\footnotesize Hellings and Downs \cite{hellingsdowns} first 
showed that for general relativity, an isotropic stochastic 
background of GWs is expected to produce the correlation shown in blue. 
The correlation for the transverse breathing mode appears in black.}
	\label{fig: HDcurve}
\end{figure}
The overlap reduction function has a closed analytic form for 
transverse GWs. The overlap reduction function for the plus mode 
has been calculated by \cite{hellingsdowns} and is given by
\begin{eqnarray}
	\Gamma_+ (\xi) = 3 \left[\frac{1}{3} +  \frac{1-\cos{\xi}}{2} 
\left[ \log{\left(\frac{1-\cos{\xi}}{2} \right)}-\frac{1}{6} \right] \right],
\end{eqnarray}
where $\xi$ is the angular separation of the pulsars. 
For the scalar-breathing mode, a closed form is given by~\cite{KJ}:
\begin{eqnarray}
	\Gamma_b (\xi) = \frac{1}{4} \left( 3+ \cos{\xi} \right).
\end{eqnarray} 
For the case of non-transverse GWs, the overlap reduction functions cannot 
be integrated analytically and we calculate them numerically.
\begin{figure*}[!]
\begin{minipage}[b]{0.45\textwidth}
\centering
	\includegraphics[width=\textwidth]{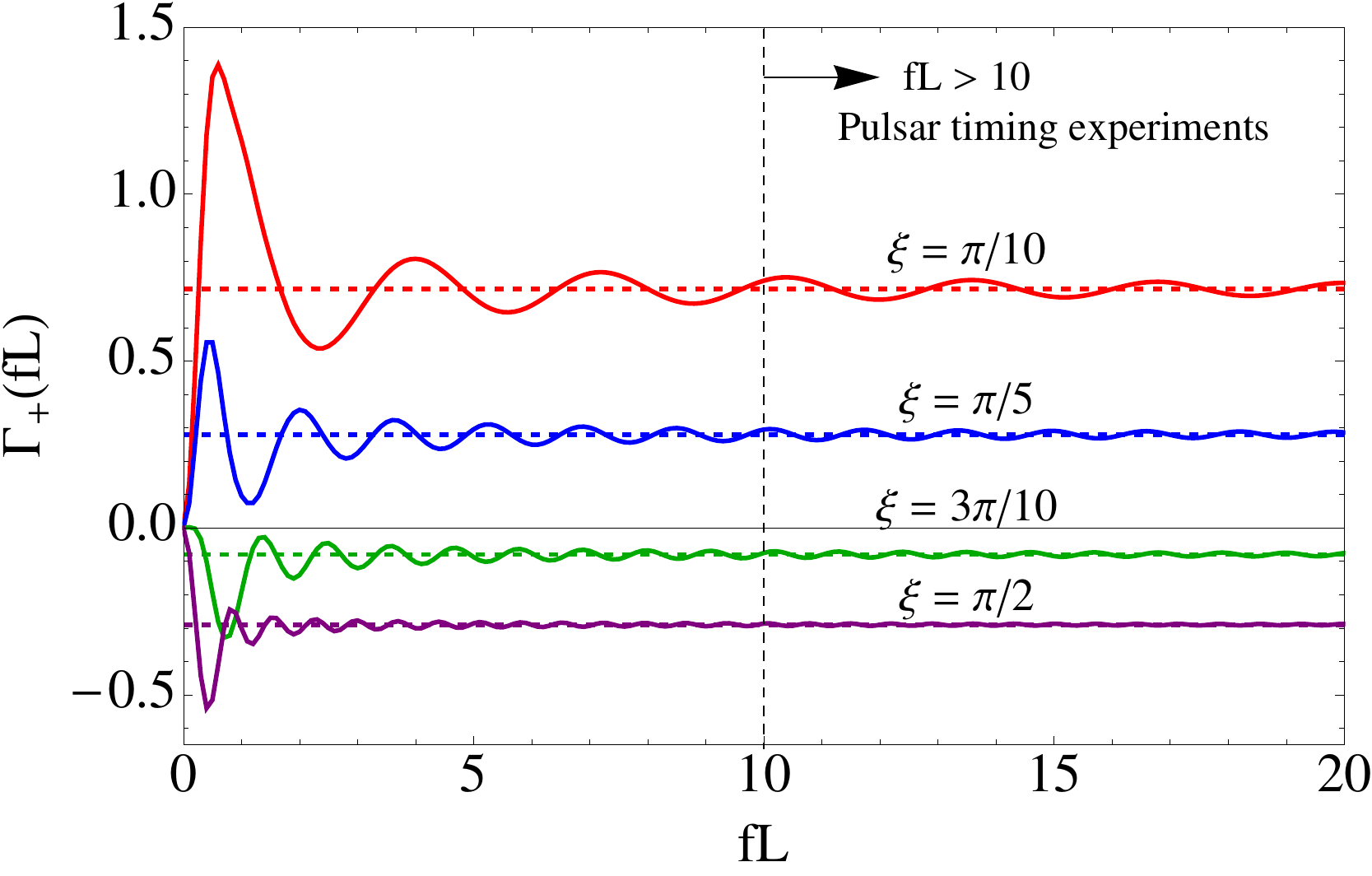}
	\centerline{(a)}
\end{minipage}
~
~
\begin{minipage}[b]{0.45\textwidth}
\centering
	\includegraphics[width=\textwidth]{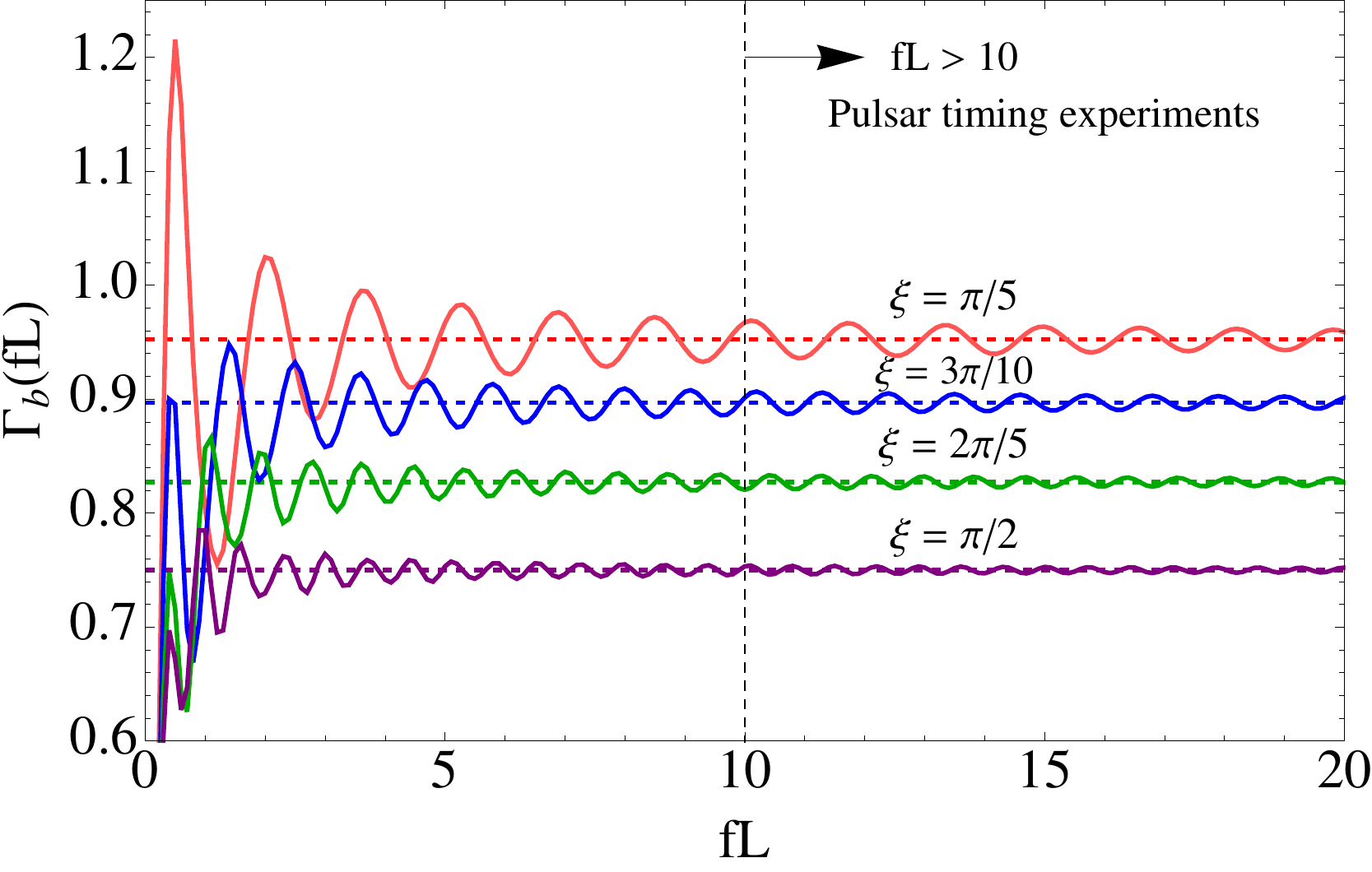}
	\centerline{(b)}
\end{minipage}
 
 \ 
 
\begin{minipage}[b]{0.45\textwidth}
\centering
	\includegraphics[width=\textwidth]{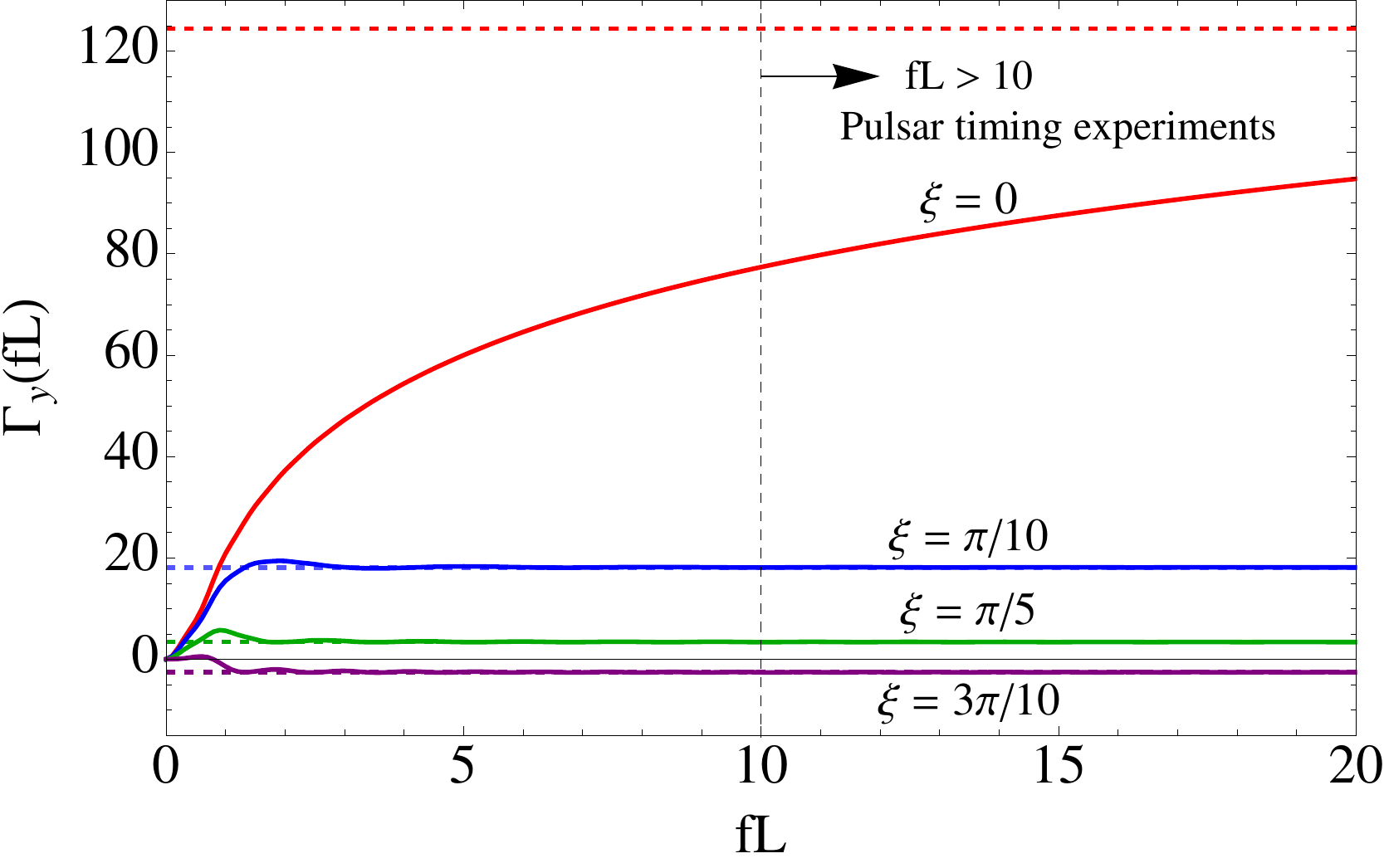}
	\centerline{(c)}
\end{minipage}
~
~	
\begin{minipage}[b]{0.45\textwidth}
\centering
	\includegraphics[width=\textwidth]{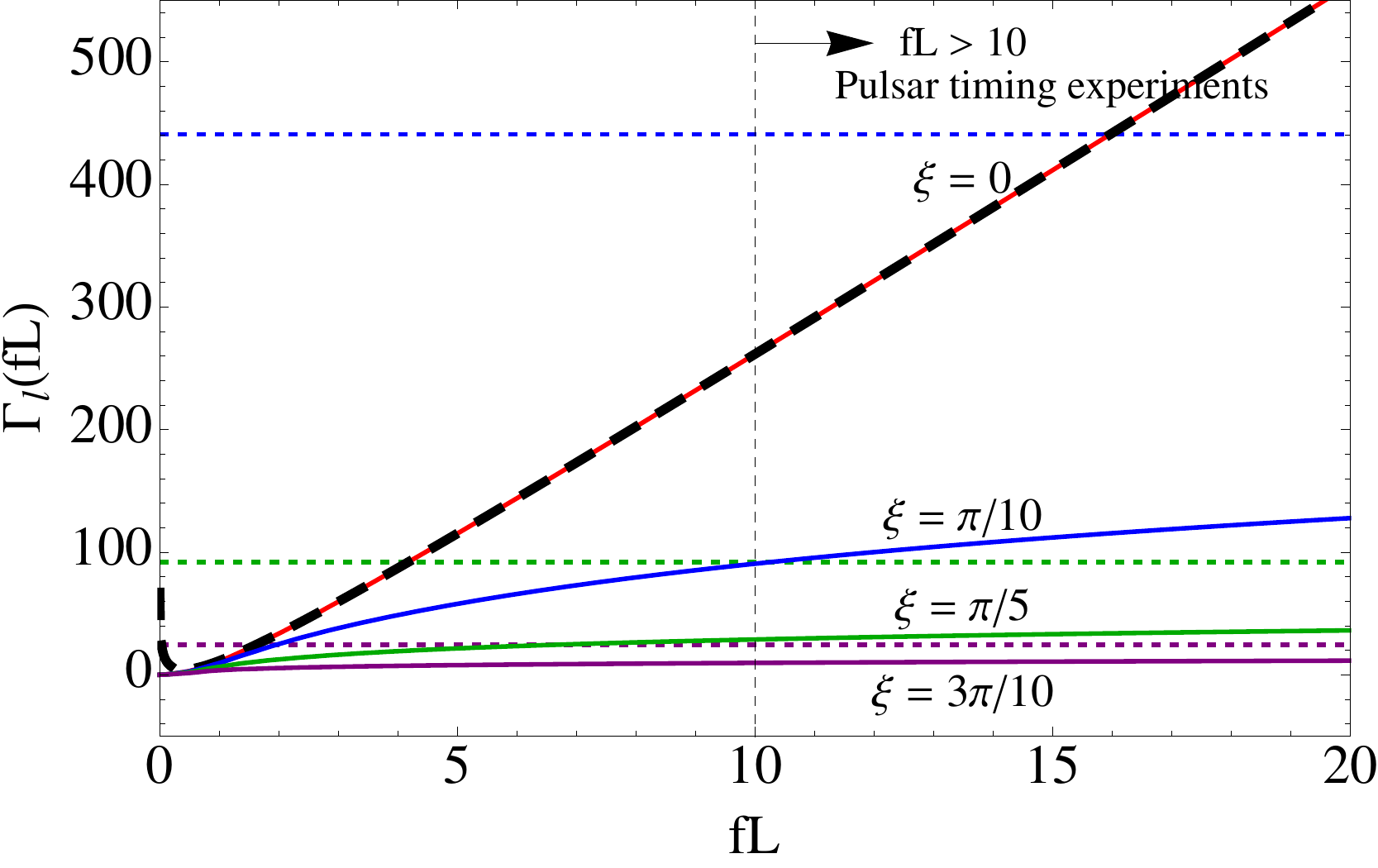}
	\centerline{(d)}
\end{minipage}
\caption{\footnotesize (color online) $\Gamma(fL)$ with (solid curves) and without (horizontal dashed lines) 
	the pulsar term for the various polarization modes: 
plus (a), breathing (b), shear (c) and longitudinal (d). In the latter 
two modes, smaller pulsar separation angles are characterized by retained 
frequency dependence in $\Gamma(fL)$ in the range of frequencies relevant to pulsar 
timing experiments. Nearly all the non-transverse curves eventually converge, but at rather high 
values of $\Gamma(fL)$ relative to the transverse modes, indicating 
increased sensitivity to GWs with these polarizations. We have plotted the 
large limit approximation \eqref{fLdep} (dashed black curve) along with $\Gamma_l(fL)$ 
in (d), which is in good agreement with the $\xi = 0$ curve.}\label{fig: an-orfs}
\end{figure*}

%

In general relativity the pulsar term can be excluded from the 
integral~\eqref{orfs} without any significant loss of optimality~\cite{anholm}. The reason for 
this is that the smallest frequencies 
that PTAs are sensitive to are $\sim 0.1$ yr$^{-1}$, and the closest PTA pulsar 
distances are $\sim 100$ ly, so that $fL\gtrsim10$. This is shown in Fig.~\ref{fig: an-orfs}, 
where we plot the overlap reduction functions $\Gamma(fL)$ with (solid curves) and without 
(horizontal dashed lines) the pulsar term for several pulsar separation angles $\xi$ 
and GW polarization modes. The frequencies that PTAs are sensitive to are to the right of 
the vertical dashed line at $fL = 10$ in each plot.
As seen in Fig.~\ref{fig: an-orfs}(a), $\Gamma_+(fL)$ is roughly independent of 
frequency over the range of frequencies relevant to pulsar timing experiments.
The same is true for the scalar-breathing mode, which is shown in 
Fig.~\ref{fig: an-orfs}(b). It is worth pointing out that both $\Gamma_{+}(fL)$ 
and $\Gamma_{b}(fL)$ are normalized to unity for co-aligned pulsars. Note that the 
overlap reduction functions
for all other modes are normalized with the same factor of $3/4\pi$ used in the $+$-mode.

In Fig.~\ref{fig: an-orfs}(c), we plot the overlap reduction function 
$\Gamma_{y}(fL)$ for the vector-y mode. Over the range of relevant frequencies,
$\Gamma_{y}(fL)$ is frequency independent for most of the pulsar separation 
angles shown. For co-aligned pulsars, however, $\Gamma_{y}(fL)$ retains 
frequency dependence well into the range of pulsar timing frequencies, 
and takes on values an order of magnitude higher than those obtained by 
$\Gamma_+(fL)$ and $\Gamma_b(fL)$.

Similar behavior is shown in Fig.~\ref{fig: an-orfs}(d), where we have 
plotted the overlap reduction function for the scalar-longitudinal 
mode. Here $\Gamma_l(fL)$ retains 
frequency dependence throughout the relevant frequency range for each 
of the pulsar separation angles shown. For the case of co-aligned 
pulsars, $\Gamma_l(fL)$ does not converge, and for separation angles 
that do converge $\Gamma_l(fL)$ takes on values that are at least an order of 
magnitude larger than those obtained by $\Gamma_+(fL)$ and $\Gamma_b(fL)$. 

For co-located pulsars we can understand the behavior of the longitudinal mode analytically. 
In the problematic sky region ($\theta \approx \pi$), 
$\Gamma_l(fL)$ is proportional to the square of the redshift,
\begin{eqnarray}
 \Gamma_l(fL) &\propto& 2 \pi \int_{-1}^1 \left\vert 
\left( e^{-2 \pi i f L (1+\cos{\theta})} -1\right)\right\vert^2 \nonumber \\
 &\times& \frac{\cos^4{\theta}}{4 (1+\cos{\theta})^2} \hspace{1mm} d{(\cos{\theta})}
\end{eqnarray}
which may be evaluated analytically. In the limit of large $fL$, 
\begin{eqnarray}\label{fLdep}
 \Gamma_l(fL) &=& \pi  \left \{37/6 - 4 \gamma -1/(\pi (fL)^2)
+ 4 \text{ Ci}({4 \pi f L}) \right. \nonumber \\
 &-& \left.   4 \log{(4 \pi fL)} +2 \pi f L\text{ Si}({4 \pi f L})\right \} 
\text{  ($fL \gg 1$)} \nonumber \\
 &\sim& \left(37/6-4 \gamma \right) \pi - 4 \pi \log{(4 \pi f L)} +  {\pi}^3 f L,
 \end{eqnarray}
where $\gamma$ is Euler's constant.
The overlap reduction function $\Gamma_l(fL)$ is roughly proportional to $f L$ in this limit.  
Eq.~\eqref{fLdep} is shown along with the numerically integrated 
overlap reduction functions in Fig. \ref{fig: an-orfs}(d) and, 
with the exception of the singular behavior near the origin (where the large $fL$ 
approximation is not valid), agrees well with 
the numerical $\Gamma_l(fL)$ curve for co-aligned pulsars ($\xi~=~0$). 

\section{Overlap reduction functions for the NANOGrav pulsars}\label{sec: NGpulsar-ORFs}

\begin{figure*}
\begin{minipage}[b]{0.45\textwidth}
\centering
	\includegraphics[width=3.4in]{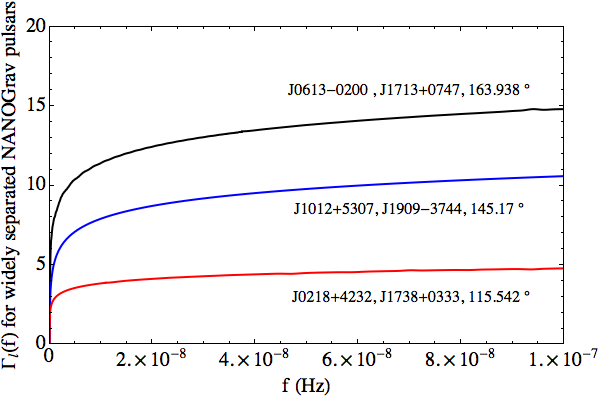}
	\centerline{(a)}
\end{minipage}
~
~
\begin{minipage}[b]{0.45\textwidth}
\centering	
	\includegraphics[width=3.5in]{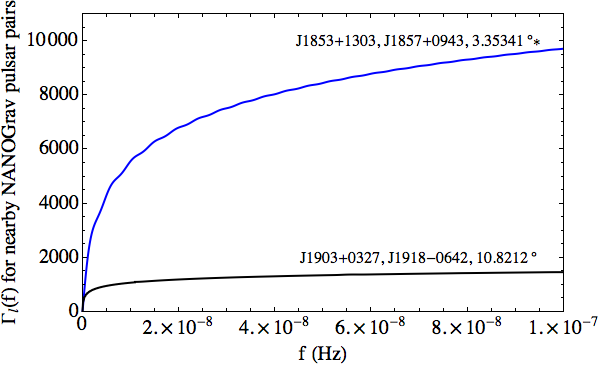}
	\centerline{(b)}
\end{minipage}

\
	
\begin{minipage}[b]{0.45\textwidth}
\centering	
	\includegraphics[width=3.5in]{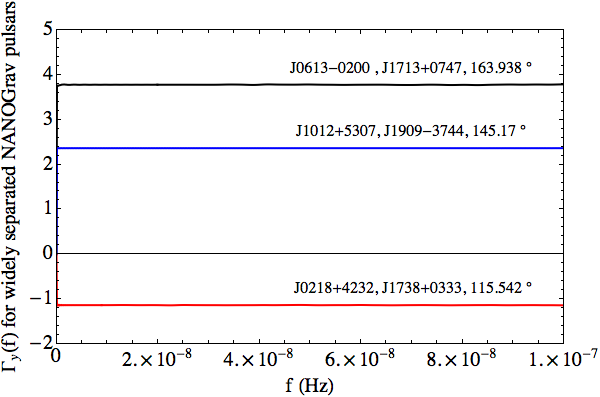}
	\centerline{(c)}
\end{minipage}
~
~	
\begin{minipage}[b]{0.45\textwidth}
\centering	
	\includegraphics[width=3.5in]{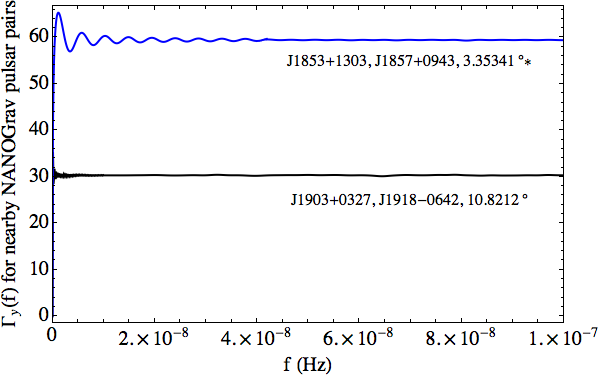}
	\centerline{(d)}
\end{minipage}

	\caption{\footnotesize $\Gamma(f)$ for some of the NANOGrav pulsar pairs. Pulsar 
pairs, along with their angular separation in degrees, are shown with each curve. As $f$ increases, $\Gamma(f)$ approaches a constant value. The 
asterisk indicates the NANOGrav pulsar pair with the smallest angular 
separation ($\sim 3.35$ degrees). Note the larger values of the $\Gamma(f)$s 
for this pair. }\label{fig: ngplot} 
\end{figure*}

The NANOGrav PTA consists of 24 pulsars. The Australia Telescope National 
Facility (ATNF) data for the distances to these pulsars is given in 
Table \ref{table: psrjs}~\cite{manchester-cat}. Using a simple numerical 
integration scheme, the overlap reduction function for each pulsar 
pair was computed. The main difference relative to the previous section is that
we are including the effect of different pulsar distances.
Results are given in Fig.~\ref{fig: ngplot} (a)--(d) 
and show that the calculated values of $\Gamma(f)$ are consistent 
with the more simple results discussed in Section \ref{sec: ORFstuff} 
for the non-transverse modes for frequencies up to $\sim 10^{-9}$ Hz.
\begin{table}
	\centering      
	\begin{tabular}{c c c | c c c} 
	\hline\hline                   
	PSR & Distance (kpc) & & & PSR & Distance (kpc) \\ 
	\hline
	J0030$+$0451 & 0.23  &  & & J1853$+$1303 & 1.60 \\
	J0218$+$4232 & 5.85  & & & J1857$+$0943 & 0.70\\
	J0613$-$0200 & 2.19   & & &  J1903$+$0327 & 6.45\\
	J1012$+$5307 & 0.52  & & & J1909$-$3744 & 0.55\\
	J1024$-$0719 & 0.35  & & & J1910$+$1256 & 1.95\\
	J1455$-$3330 & 0.74  & & & J1918$-$0642 & 1.40\\
	J1600$-$3053 & 2.67  & & & J1939$+$2134 & 3.58\\
	J1640$+$2224& 1.19 & & & J1944$+$0907 & 1.28\\
	J1643$-$1224 & 4.86 & & & J1955$+$2908 & 5.39\\
	J1713$+$0747 & 0.89 & & & J2010$-$1323 & 1.29\\
	J1738$+$0333 & 1.97 & & & J2145$-$0750 & 0.50\\
	J1744$-$1134 & 0.17 & & & J2317$+$1439 & 1.89\\
	\hline\hline
	\end{tabular}  \caption{\footnotesize NANOGrav Pulsar Data}\label{table: psrjs}
\end{table} 
Pulsar pairs with the smallest ($\xi \lesssim 12^{\circ}$) separation 
angles (starred curves in Fig.~\ref{fig: ngplot} (b), (d)) for non-transverse
polarization modes are characterized 
by large values of the overlap reduction function and monotonic growth 
up to some limiting frequency. Pulsar pairs with larger ($\xi \gtrsim 12^{\circ}$) 
separation angles (un-starred curves in Fig.~\ref{fig: ngplot} (b), (d) and 
all curves in Fig.~\ref{fig: ngplot}) do not display monotonic growth up 
to a limiting frequency, but still result in much larger values than those 
of the plus and cross modes. Fig.~\ref{fig: ngplot} shows that sensitivity 
is greater for scalar-longitudinal and vector modes than for the tensor 
and scalar-breathing modes, and increases rapidly for pulsars that 
are nearly co-aligned in the sky. 

Over the entire range of frequencies plotted for pulsar timing experiments
(between $\sim 10^{-9}$ and $\sim 10^{-7}$~Hz), the overlap reduction functions 
are approximately constant. In practice, some optimality will be lost due to 
the fact that pulsar distances are known at best to 
only $\sim10\%$~\cite{cordes-lazio}. 
%

%
\section{Discussion}\label{discussion}
Direct detection of GWs might be possible in the next decade using a 
pulsar timing array. A detection would provide a mechanism for 
testing various metric theories of gravity. To develop optimal
detection strategies for stochastic backgrounds in alternative 
theories of gravity, we have computed overlap reduction functions 
for all six GW polarization modes, including four modes not present in general 
relativity. 

We began by introducing the 
redshift induced by GWs of various polarizations, along with the 
polarization tensors unique to each mode. We then used the optimal 
detection statistic for an unpolarized, isotropic stochastic background of 
GWs, defined in Anholm et al. \cite{anholm}, to find the overlap reduction 
function, a geometric dependent quantity in the expression for the 
expected cross correlation.

We examined the redshifts induced by GWs 
of various polarizations on the pulsar-Earth system, and find that our results
are consistent with those of Anholm et al. \cite{anholm} and Tinto and Alves 
\cite{tintoalves2010}: when the GWs are coming from
roughly the same direction as the pulses from the pulsar, the induced 
redshift for \emph{any} GW polarization mode
is proportional to $fL$, the product of the GW frequency and the distance to the pulsar. 
When the GWs and the pulse direction are exactly parallel the 
redshift for the transverse and vector modes vanishes, but it is proportional
to $fL$ for the scalar-longitudinal mode. 

We show that the vanishing contributions from the 
tensor, vector and scalar-breathing modes are not a result of the pulse 
surfing the GW. In fact, sensitivity to GWs coming from directions near 
the pulsar increases for \emph{all} polarizations. It is the transverse 
nature of these modes that is responsible for the vanishing response. In 
this limit we also show that the redshift is proportional to the relative 
velocity of the pulsar-Earth system ($L\dot{h}$), which is the same 
when the pulse is emitted and when it is received. 

We find that the overlap reduction functions for non-transverse GWs 
are characterized by frequency dependence that is significant for 
nearby pulsar pairs. The values of the
overlap reduction function increase by up to one order
of magnitude for the vector polarization modes and up to two
orders of magnitude for the scalar-longitudinal 
mode. Pulsar timing arrays are significantly 
more sensitive to scalar-longitudinal and 
vector GW stochastic backgrounds. 

Next, we used current pulsar distance and sky-location 
data from the ATNF pulsar catalog
to calculate the overlap reduction functions for each pulsar pair 
in the NANOGrav pulsar timing array. Over the range of frequencies 
relevant to pulsar timing array experiments, these overlap reduction 
functions for all polarization modes 
are roughly constant for most pulsar pairs. For nearly co-aligned pulsars, 
the overlap reduction functions for scalar-longitudinal and vector modes
exhibit marked frequency dependence 
and asymptote to much larger values than the overlap reduction functions 
for transverse modes. In fact for a pair separated but about $3^\circ$ 
we find a sensitivity increase of about a factor of $10^4$ for longitudinal modes.

The results discussed here may be compared to other recent work. 
Lee et al. \cite{KJ} calculated the cross-correlation functions for 
stochastic GW backgrounds including all six GW polarizations, and found
that the correlation functions for non-transverse GWs are frequency dependent,
as well as an increased response in the cross-correlation to scalar-longitudinal
GWs, in agreement with our results. This work was done in the context of the 
coherence statistic~\cite{KJ} for stochastic background detection, 
rather than the optimal statistic~\cite{anholm}.  The coherence statistic 
is a measure of goodness of fit of the pulsar-pair cross-correlations to 
the Hellings-Downs curve. For non-transverse modes there is no Hellings-Downs curve
because the overlap reduction functions remain frequency dependent for large $fL$. 
Lee et al. solved this 
problem by simulating GW backgrounds and finding effective background-dependent 
Hellings-Downs curves for these theories. In the context of the optimal statistic this is a 
non-issue: The frequency dependent overlap reduction functions can be used 
to construct the optimal filter in Eq.~\eqref{optimalfilter}. 
This is identical to what is done for LIGO stochastic background optimal 
filter construction~\cite{allenromano}, where the overlap reduction functions 
are also frequency dependent.
 
Alves and Tinto \cite{alvestinto} have estimated antenna
sensitivities to GWs of all six polarization modes by assuming 
a signal-to-noise ratio of 1 over 10 years and calculating the 
noise spectrum. Their results indicate an increase of two
to three orders of magnitude in sensitivity to scalar-longitudinal mode GWs
compared to that of plus and cross mode GWs. To explain this effect Alves and 
Tinto compare the effect of a tensor GW propagating orthogonally 
to the pulsar-Earth system, and a scalar-longitudinal GW propagating in a direction 
parallel to the pulse direction. They argue that the increased sensitivity 
to longitudinal GWs is due to the amount of time a 
longitudinal GW affects the pulsar-Earth radio link. 

We have compared the effect of GW propagation from directions near the pulsar and 
orthogonal to the pulsar-Earth system for all
polarization modes. 
For GW propagation directions parallel to the pulse direction
we find that the redshift induced by a gravitational wave is large, 
and seemingly divergent 
when the GW and pulse directions are exactly parallel.  This apparent divergence 
occurs for longitudinal, transverse, and shear modes alike. 
In that limit, however, the divergent term 
in the redshift that comes from the relationship between time and affine parameter 
derivatives cancels because the phase of the GW pulse when pulse is emitted is nearly equal
to the phase of the GW when the pulse is received (see 
Eqs.~\eqref{affineparam}, \eqref{gen-redshift} and \eqref{expanded-zA}). 
The redshift becomes proportional to the relative velocity 
of the pulsar-Earth system and a mode-dependent geometrical projection factor  
for {\it all} GW polarization modes. 
In this limit the relative velocity of the pulsar-Earth system
is approximately equal when the pulse is emitted and
received. For transverse and shear modes the projection 
factor vanishes when the GW and pulse directions become parallel. For longitudinal 
modes the geometrical factor goes to a constant, so that the pulsar-Earth 
system is very sensitive 
to GWs from directions near the pulsar. This is the physical origin of the 
increased sensitivity to scalar-longitudinal GWs.

\begin{acknowledgements}
We would like to extend our thanks to the members of the NANOGrav data analysis 
working group for their comments and support, especially Jim Cordes, 
Paul Demorest, Justin Ellis, Rick Jenet,
Andrea Lommen, Delphine Perrodin, Sam Finn, and Joe Romano. We would also like to thank 
Jolien D. E. Creighton for numerous useful comments and suggestions. This 
work was funded in part by the Wisconsin Space Grant 
Consortium and the NSF through CAREER award number 09955929 and PIRE award 
number 0968126.
\end{acknowledgements}

\appendix
\section{Analog to Detweiler's equation for vector and scalar polarization modes}\label{sec: appA}

Here we show the derivation of the redshift induced by non-Einsteinian GW modes. This derivation 
appears in~\cite{tintoalves2010} for all six GW polarizations and is included here for completeness.
We begin by considering the metric due to a longitudinal mode gravitational wave perturbation:
\begin{eqnarray}
g_{ab}&=&\eta_{ab} + h_{ab}(t-z) \nonumber \\
&=&\left( \begin{array}{cccc} -1 & 0 & 0 & 0\\ 0 &1 & 0 & 0 \\ 0 & 0 & 1 & 0 \\ 0 & 0 & 0 & 1 + h_L 
\end{array} \right).
\end{eqnarray}
Given a null vector $s^a = \nu (1, -\alpha, -\beta,-\gamma)$ in Minkowski 
space (where $\alpha\text{, }\beta\text{, }\gamma$ are directional cosines) 
the corresponding perturbed null vector is given by 
\begin{eqnarray}
	\sigma^a &=& s^a - \frac{1}{2} \eta^{ab} h_{bc} s^c \nonumber \\
	&=& \nu \left( \begin{array}{c} 1 \\ -\alpha \\ -\beta \\ -\gamma 
(1-\frac{h_L}{2}) \end{array} \right) .
\end{eqnarray}
From the geodesic equation, the t-component of $\sigma^a$ must satisfy
\begin{eqnarray}
	\frac{d\sigma^t}{d\lambda} &=& -\Gamma^t_{ab} \sigma^a \sigma^b
\end{eqnarray}
where 
\begin{eqnarray}
	\Gamma^t_{ab} &=& \frac{1}{2} g^{tc} \left( \partial_a g_{bc} + 
\partial_b g_{ac} - \partial_c g_{ab} \right) \nonumber\\
	&=& \frac{1}{2} \dot{g}_{ab}.
\end{eqnarray}
Now we may write the geodesic equation as 
\begin{eqnarray}
	\frac{d\sigma^t}{d\lambda} &=& -\frac{1}{2} 
\dot{g}_{ab} \sigma^a \sigma^b \nonumber \\
	&=& -\frac{1}{2} \dot{h_L} (\sigma^z)^2.
\end{eqnarray}
To zeroth order in $h_L$, 
\begin{eqnarray}
	(\sigma^z)^2 &=&\nu^2 \gamma^2 \left( 1+ \frac{h_L}{2} \right)^2 \nonumber \\
	&\approx&\nu^2 \gamma^2 + O(h_L)
\end{eqnarray}
allowing us to write the geodesic equation as  
\begin{eqnarray}
	\frac{d\sigma^t}{d\lambda} &=& \frac{d\nu}{d\lambda} 
= -\frac{1}{2} \dot{h_L}\nu^2 \gamma^2.
\end{eqnarray}
We now need to express the time derivative of the metric perturbation, $\dot{h_L}$, 
as a derivative of the affine parameter $\lambda$. Since $h_L = h_L (t-z)$, we may write
\begin{eqnarray}
	\frac{d h_L}{d\lambda} &=& \frac{\partial h_L}{\partial t} \frac{d t}{d \lambda} 
+  \frac{\partial h_L}{\partial z} \frac{d z}{d \lambda} \nonumber \\
	&=&  \frac{\partial h_L}{\partial t} \frac{d t}{d \lambda} 
-  \frac{\partial h_L}{\partial t} \frac{d z}{d \lambda}.
\end{eqnarray}
Identifying the relations $\frac{dt}{d\lambda} = \nu$ and $\frac{dz}{d\lambda} = -\nu \gamma$, 
we obtain the relation
\begin{eqnarray}\label{affineparam}
	\dot{h_L} = \frac{\partial h_L}{ \partial t} = \frac{1}{\nu (1+\gamma)}\frac{dh_L}{d\lambda}
\end{eqnarray}
which makes the geodesic equation 
\begin{eqnarray}
	\frac{d\nu}{d\lambda} &=& -\frac{1}{2} \dot{h_L} \nu^2 \gamma^2 = 
-\frac{1}{2} \frac{\nu \gamma^2}{ (1+\gamma)} \frac{dh_L}{d\lambda} 
\end{eqnarray}
Integrating both sides, we obtain
\begin{eqnarray}
	\frac{\nu_e}{\nu_p} = \exp{\left(-\frac{1}{2} \frac{\gamma^2}
{(1+\gamma)} \Delta h_L \right)}
\end{eqnarray}
where $\Delta h_l = h^e_l - h^p_l$. Expanding to first order in $h_L$, we may write
\begin{eqnarray}\label{longz}
	\frac{\nu_e-\nu_p}{\nu_p} &\approx& -\frac{1}{2} 
\frac{\gamma^2}{(1+\gamma)} \Delta h_L \\
&=& - \frac{\cos^2{\theta_p}}{2\left(1+\cos{\theta_p} \right)}  \Delta h_L.
\end{eqnarray}
The derivation for vector modes is nearly identical to that of the 
longitudinal mode. For the sake of brevity we only detail the 
vector-y mode in the remainder of this document. For the vector-y mode, the metric 
perturbation takes the form 
\begin{eqnarray}
g_{ab}=\left( \begin{array}{cccc} -1 & 0 & 0 & 0\\ 0 &1 & 0 & 0 \\ 
0 & 0 & 1 & h_y \\ 0 & 0 & h_y & 1 \end{array} \right).
\end{eqnarray}
The null vector becomes 
\begin{eqnarray}
	\sigma^a = \nu \left( \begin{array}{c} 1 \\ -\alpha \\ 
-\beta+ \frac{h_y \gamma}{2} \\ \frac{h_y \beta}{2} - \gamma \end{array} \right) .
\end{eqnarray}
Following the same algebraic steps used above, one obtains the geodesic equation 
\begin{eqnarray}
	\frac{d\sigma^t}{d\lambda} &=& \frac{d\nu}{d\lambda} = - \dot{h_y}\nu^2 \gamma \beta,
\end{eqnarray}
which leads to 
\begin{eqnarray}
	\frac{d\nu}{d\lambda} = -\frac{\nu \gamma \beta }{(1+\gamma)} 
\frac{d h_y}{d \lambda}.
\end{eqnarray}
Integrating this expression and expanding the result to first order in 
$\Delta h_y$ produces the result
\begin{eqnarray}\label{yz}
	\frac{\nu_e - \nu_p}{\nu_p} &\approx& - \frac{\beta \gamma}{(1+\gamma)} \Delta h_y\\
	&=& - \frac{\sin{2 \theta_p} \sin{\phi_p}}{2 \left(1+\cos{\theta_p} \right)}  \Delta h_y.
\end{eqnarray}
where $\Delta h_y = h^e_y - h^p_y$.

For comparison, we also include the results for the plus, cross, vector-x, and breathing modes. 
For the plus mode, we obtain
\begin{eqnarray}\label{plusz}
	\frac{\nu_e - \nu_p}{\nu_p} &\approx& - \frac{\left( \alpha^2-\beta^2 \right)}{2(1+\gamma)} \Delta h_+\\
	&=& -\frac{\sin^2{\theta_p} \cos{2 \phi_p}}{2 \left(1+\cos{\theta_p} \right)} \Delta h_+;
\nonumber
\\
\end{eqnarray}
for the cross mode, 
\begin{eqnarray}\label{crossz}
	\frac{\nu_e - \nu_p}{\nu_p} &\approx& - \frac{\alpha \beta}{(1+\gamma)} \Delta h_{\times} \\
	&=& -\frac{\sin^2{\theta_p} \sin{2 \phi_p}}{2 \left(1+\cos{\theta_p} \right)}  \Delta h_{\times};
\end{eqnarray}
for the vector-x mode,
\begin{eqnarray}\label{xz}
	\frac{\nu_e - \nu_p}{\nu_p} &\approx& - \frac{\alpha \gamma }{(1+\gamma)} \Delta h_x\\
	&=& -\frac{\sin{2 \theta_p} \cos{\phi_p}}{2 \left(1+\cos{\theta_p} \right)} \Delta h_x;
\end{eqnarray}
 and for the breathing mode, 
\begin{eqnarray}\label{breathez}
	\frac{\nu_e - \nu_p}{\nu_p} &\approx& -\frac{\left( \alpha^2 + \beta^2 \right) }{2(1+\gamma)} \Delta h_b\\
	&=& \frac{-\sin^2{\theta_p}}{2 \left(1+\cos{\theta_p} \right)}\Delta h_b.
\nonumber
\\
\end{eqnarray}
Here, $\Delta h_A = h_A^e - h_A^p$, and we can identify these expressions with \eqref{reddef}.   

\begin{figure}[t]
	\includegraphics[width=2.5in]{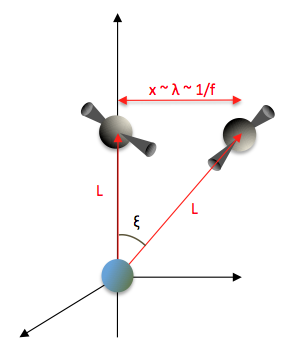}
\caption{\footnotesize A system of two pulsars, distance $L$ form the Earth, are shown along with their separation angle $\xi$ and separation distance $x \approx L \xi$. When the GW is in the long wavelength limit, this separation distance is proportional to the GW wavelength. }\label{fig: appBdiagram}
\end{figure} 

\begin{figure}[t]
	\includegraphics[width=3.5in]{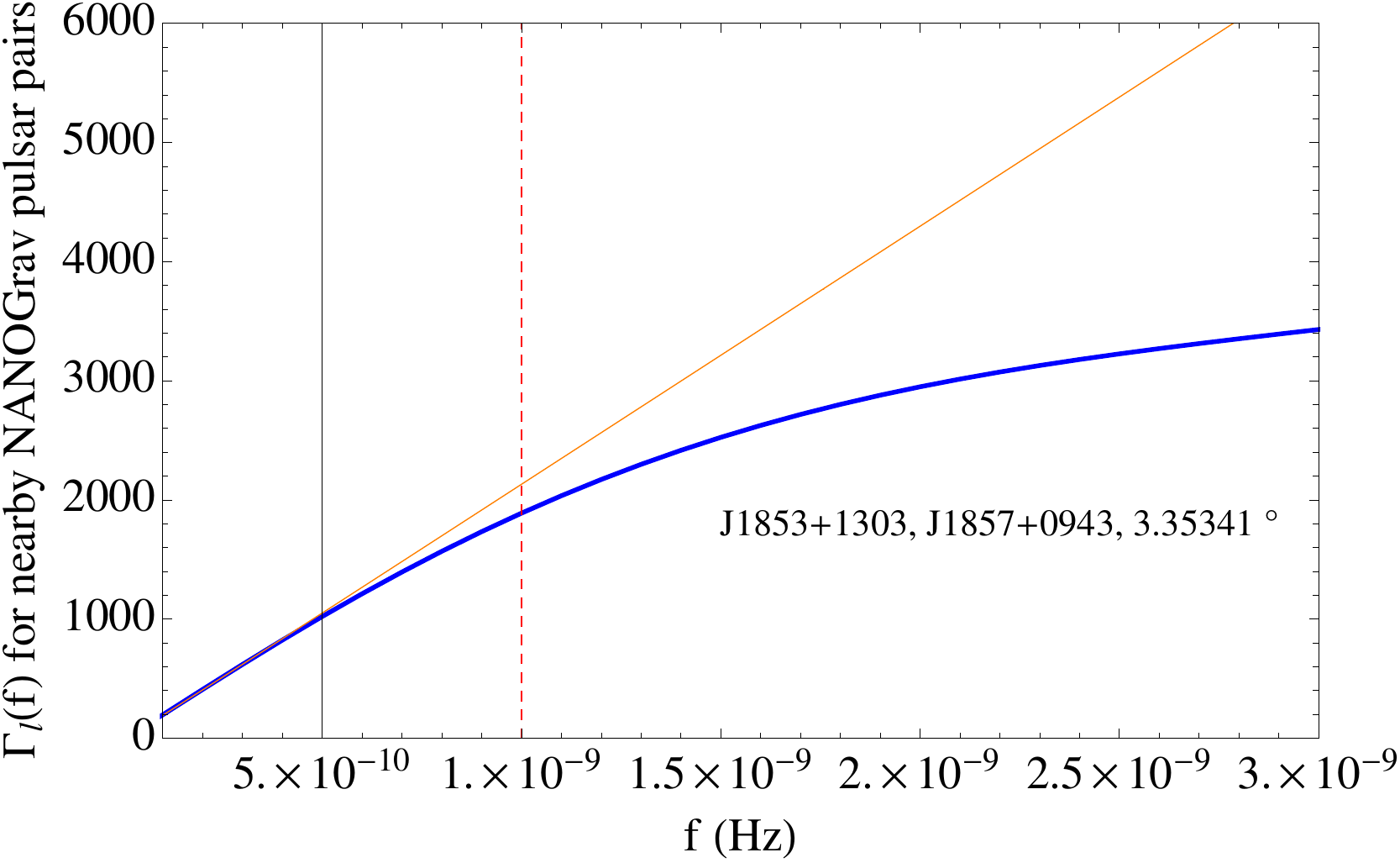}
	\caption{\footnotesize The NANOGrav pulsar pair $J1853+1303$, $J1857+0943$ has an angular separation of nearly $3^\circ$, with each pulsar approximately $1$ kpc from the Earth. Using the estimate \eqref{regime-end-fold}, the curve $\Gamma_l(f)$ should stop growing as $\sim \pi^3 fL$ near $10^{-9}$ Hz, which is shown as the red vertical dashed line. Note, however, that the curve does not converge onto constant values of $\Gamma_l(f)$ at this point; in fact the overlap reduction function continues to grow well past this point, but no longer linearly with $fL$ (as indicated by orange solid line).}\label{fig: zoomedNG} 
\end{figure}
$\,$
\section{Trends in $\Gamma_l(f)$ for nearby pulsar pairs }\label{appB}

Consider a pair of pulsars separated by some \emph{small} angle $\xi$
and located approximately equidistant from the Earth so that $L_1 \sim
L_2 \equiv L$. 

As shown in Section~\ref{sec: ORFstuff},  if the two pulsars are
co-located the overlap reduction function 
$\Gamma_l(f)\sim \pi^3 f L$.  We expect that if they are separated by
a small angle the overlap reduction function will increase as $\pi^3 f
L$ as though they were co-located,
until the wavelength of the GW is comparable to the distance between
the two pulsars. This happens when
\begin{eqnarray}
  \lambda \sim L \xi,
\end{eqnarray}
so that the value of $fL$ where the behavior changes from the
co-located case is
\begin{eqnarray}\label{regime-end-fold}
	fL \sim 1/\xi.
\end{eqnarray}
For example, for the closest NANOGrav pulsar pair, separated  by an angle $\xi \sim
3^\circ$ at a distance of $L \sim 1$~kpc, the frequency at which the linear growth of the overlap reduction
function stops is
\begin{eqnarray}
	f \sim 10^{-9} \text{ Hz}.
\end{eqnarray}
The value of of the overlap reduction function where the behavior
changes from the co-located case  $\Gamma_l(\xi^{-1})$ is a poor estimate of
the maximum value of $\Gamma_l(f)$, however, because after
exiting the linear regime of Eq.~\eqref{fLdep}, the overlap reduction functions
continue to increase significantly before converging. 


A closer look at the lower frequency portion of the plot Fig.~\ref{fig: ngplot}(b), shown in Fig.~\ref{fig: zoomedNG}, indicates that to order of magnitude this approximation is roughly valid.

\bibliographystyle{apsrev}	
\bibliography{researchbiblio}

\begin{thebibliography}{30}
\expandafter\ifx\csname natexlab\endcsname\relax\def\natexlab#1{#1}\fi
\expandafter\ifx\csname bibnamefont\endcsname\relax
  \def\bibnamefont#1{#1}\fi
\expandafter\ifx\csname bibfnamefont\endcsname\relax
  \def\bibfnamefont#1{#1}\fi
\expandafter\ifx\csname citenamefont\endcsname\relax
  \def\citenamefont#1{#1}\fi
\expandafter\ifx\csname url\endcsname\relax
  \def\url#1{\texttt{#1}}\fi
\expandafter\ifx\csname urlprefix\endcsname\relax\def\urlprefix{URL }\fi
\providecommand{\bibinfo}[2]{#2}
\providecommand{\eprint}[2][]{\url{#2}}

\bibitem[{\citenamefont{{Nojiri} and {Odintsov}}(2006)}]{noj-odi}
\bibinfo{author}{\bibfnamefont{S.}~\bibnamefont{{Nojiri}}} \bibnamefont{and}
  \bibinfo{author}{\bibfnamefont{S.~D.} \bibnamefont{{Odintsov}}},
  \bibinfo{journal}{ArXiv High Energy Physics - Theory e-prints}
  (\bibinfo{year}{2006}), \eprint{arXiv:hep-th/0601213}.

\bibitem[{\citenamefont{{Nojiri} and {Odintsov}}(2011)}]{noj-odi-newer}
\bibinfo{author}{\bibfnamefont{S.}~\bibnamefont{{Nojiri}}} \bibnamefont{and}
  \bibinfo{author}{\bibfnamefont{S.~D.} \bibnamefont{{Odintsov}}},
  \bibinfo{journal}{Phys. Rep.} \textbf{\bibinfo{volume}{505}},
  \bibinfo{pages}{59} (\bibinfo{year}{2011}), \eprint{1011.0544}.

\bibitem[{\citenamefont{{Lobo}}(2008)}]{Lobo:2008sg}
\bibinfo{author}{\bibfnamefont{F.~S.~N.} \bibnamefont{{Lobo}}},
  \bibinfo{journal}{ArXiv e-prints}  (\bibinfo{year}{2008}),
  \eprint{0807.1640}.

\bibitem[{\citenamefont{{Alves} et~al.}(2009)\citenamefont{{Alves}, {Miranda},
  and {de Araujo}}}]{Alves:2009eg}
\bibinfo{author}{\bibfnamefont{M.~E.~S.} \bibnamefont{{Alves}}},
  \bibinfo{author}{\bibfnamefont{O.~D.} \bibnamefont{{Miranda}}},
  \bibnamefont{and} \bibinfo{author}{\bibfnamefont{J.~C.~N.} \bibnamefont{{de
  Araujo}}}, \bibinfo{journal}{Phys. Lett. B} \textbf{\bibinfo{volume}{679}},
  \bibinfo{pages}{401} (\bibinfo{year}{2009}), \eprint{0908.0861}.

\bibitem[{\citenamefont{Capozziello and
  Francaviglia}(2008)}]{Capozziello:2007ec}
\bibinfo{author}{\bibfnamefont{S.}~\bibnamefont{Capozziello}} \bibnamefont{and}
  \bibinfo{author}{\bibfnamefont{M.}~\bibnamefont{Francaviglia}},
  \bibinfo{journal}{Gen. Rel. Grav.} \textbf{\bibinfo{volume}{40}},
  \bibinfo{pages}{357} (\bibinfo{year}{2008}), \eprint{0706.1146}.

\bibitem[{\citenamefont{De~Felice and Tsujikawa}(2010)}]{DeFelice:2010aj}
\bibinfo{author}{\bibfnamefont{A.}~\bibnamefont{De~Felice}} \bibnamefont{and}
  \bibinfo{author}{\bibfnamefont{S.}~\bibnamefont{Tsujikawa}},
  \bibinfo{journal}{Living Rev. Rel.} \textbf{\bibinfo{volume}{13}}
  (\bibinfo{year}{2010}), \eprint{1002.4928}.

\bibitem[{\citenamefont{{Will}}(1993)}]{willbook}
\bibinfo{author}{\bibfnamefont{C.~M.} \bibnamefont{{Will}}},
  \emph{\bibinfo{title}{{Theory and Experiment in Gravitational Physics}}}
  (\bibinfo{publisher}{Cambridge University Press}, \bibinfo{year}{1993}).

\bibitem[{\citenamefont{Brunetti et~al.}(1999)\citenamefont{Brunetti, Coccia,
  Fafone, and Fucito}}]{brunetti-1999-59}
\bibinfo{author}{\bibfnamefont{M.}~\bibnamefont{Brunetti}},
  \bibinfo{author}{\bibfnamefont{E.}~\bibnamefont{Coccia}},
  \bibinfo{author}{\bibfnamefont{V.}~\bibnamefont{Fafone}}, \bibnamefont{and}
  \bibinfo{author}{\bibfnamefont{F.}~\bibnamefont{Fucito}},
  \bibinfo{journal}{\prd} \textbf{\bibinfo{volume}{59}},
  \bibinfo{pages}{044027} (\bibinfo{year}{1999}),
  \urlprefix\url{http://www.citebase.org/abstract?id=oai:arXiv.org:gr-qc/9805056}.

\bibitem[{\citenamefont{{Clifton} et~al.}(2011)\citenamefont{{Clifton},
  {Ferreira}, {Padilla}, and {Skordis}}}]{Clifton:2011jh}
\bibinfo{author}{\bibfnamefont{T.}~\bibnamefont{{Clifton}}},
  \bibinfo{author}{\bibfnamefont{P.~G.} \bibnamefont{{Ferreira}}},
  \bibinfo{author}{\bibfnamefont{A.}~\bibnamefont{{Padilla}}},
  \bibnamefont{and}
  \bibinfo{author}{\bibfnamefont{C.}~\bibnamefont{{Skordis}}},
  \bibinfo{journal}{ArXiv e-prints}  (\bibinfo{year}{2011}),
  \eprint{1106.2476}.

\bibitem[{\citenamefont{Sagi}(2010)}]{Sagi:2010ei}
\bibinfo{author}{\bibfnamefont{E.}~\bibnamefont{Sagi}}, \bibinfo{journal}{\prd}
  \textbf{\bibinfo{volume}{81}}, \bibinfo{pages}{064031}
  (\bibinfo{year}{2010}), \eprint{1001.1555}.

\bibitem[{\citenamefont{Clifton et~al.}(2010)\citenamefont{Clifton, Banados,
  and Skordis}}]{Clifton:2010hz}
\bibinfo{author}{\bibfnamefont{T.}~\bibnamefont{Clifton}},
  \bibinfo{author}{\bibfnamefont{M.}~\bibnamefont{Banados}}, \bibnamefont{and}
  \bibinfo{author}{\bibfnamefont{C.}~\bibnamefont{Skordis}},
  \bibinfo{journal}{Class. Quant. Grav.} \textbf{\bibinfo{volume}{27}},
  \bibinfo{pages}{235020} (\bibinfo{year}{2010}), \eprint{1006.5619}.

\bibitem[{\citenamefont{Skordis}(2009)}]{Skordis:2009bf}
\bibinfo{author}{\bibfnamefont{C.}~\bibnamefont{Skordis}},
  \bibinfo{journal}{Class. Quant. Grav.} \textbf{\bibinfo{volume}{26}},
  \bibinfo{pages}{143001} (\bibinfo{year}{2009}), \eprint{0903.3602}.

\bibitem[{\citenamefont{Milgrom}(2009)}]{Milgrom:2009gv}
\bibinfo{author}{\bibfnamefont{M.}~\bibnamefont{Milgrom}},
  \bibinfo{journal}{\prd} \textbf{\bibinfo{volume}{80}},
  \bibinfo{pages}{123536} (\bibinfo{year}{2009}), \eprint{0912.0790}.

\bibitem[{\citenamefont{Abramovici et~al.}(1992)\citenamefont{Abramovici,
  Althouse, Drever, Gursel, Kawamura et~al.}}]{Abramovici:1992ah}
\bibinfo{author}{\bibfnamefont{A.}~\bibnamefont{Abramovici}},
  \bibinfo{author}{\bibfnamefont{W.~E.} \bibnamefont{Althouse}},
  \bibinfo{author}{\bibfnamefont{R.~W.} \bibnamefont{Drever}},
  \bibinfo{author}{\bibfnamefont{Y.}~\bibnamefont{Gursel}},
  \bibinfo{author}{\bibfnamefont{S.}~\bibnamefont{Kawamura}},
  \bibnamefont{et~al.}, \bibinfo{journal}{Science}
  \textbf{\bibinfo{volume}{256}}, \bibinfo{pages}{325} (\bibinfo{year}{1992}).

\bibitem[{\citenamefont{Hobbs et~al.}(2010)\citenamefont{Hobbs, Archibald,
  Arzoumanian, Backer, Bailes et~al.}}]{Hobbs:2009yy}
\bibinfo{author}{\bibfnamefont{G.}~\bibnamefont{Hobbs}},
  \bibinfo{author}{\bibfnamefont{A.}~\bibnamefont{Archibald}},
  \bibinfo{author}{\bibfnamefont{Z.}~\bibnamefont{Arzoumanian}},
  \bibinfo{author}{\bibfnamefont{D.}~\bibnamefont{Backer}},
  \bibinfo{author}{\bibfnamefont{M.}~\bibnamefont{Bailes}},
  \bibnamefont{et~al.}, \bibinfo{journal}{Class. Quant. Grav.}
  \textbf{\bibinfo{volume}{27}}, \bibinfo{pages}{084013}
  (\bibinfo{year}{2010}), \eprint{0911.5206}.

\bibitem[{\citenamefont{{Sesana} et~al.}(2008)\citenamefont{{Sesana},
  {Vecchio}, and {Colacino}}}]{Sesana:2008mz}
\bibinfo{author}{\bibfnamefont{A.}~\bibnamefont{{Sesana}}},
  \bibinfo{author}{\bibfnamefont{A.}~\bibnamefont{{Vecchio}}},
  \bibnamefont{and} \bibinfo{author}{\bibfnamefont{C.~N.}
  \bibnamefont{{Colacino}}}, \bibinfo{journal}{Mon. Not. R. Astron. Soc.}
  \textbf{\bibinfo{volume}{390}}, \bibinfo{pages}{192} (\bibinfo{year}{2008}),
  \eprint{0804.4476}.

\bibitem[{\citenamefont{Olmez et~al.}(2010)\citenamefont{Olmez, Mandic, and
  Siemens}}]{Olmez:2010bi}
\bibinfo{author}{\bibfnamefont{S.}~\bibnamefont{Olmez}},
  \bibinfo{author}{\bibfnamefont{V.}~\bibnamefont{Mandic}}, \bibnamefont{and}
  \bibinfo{author}{\bibfnamefont{X.}~\bibnamefont{Siemens}},
  \bibinfo{journal}{\prd} \textbf{\bibinfo{volume}{81}},
  \bibinfo{pages}{104028} (\bibinfo{year}{2010}), \eprint{1004.0890}.

\bibitem[{\citenamefont{Starobinsky}(1979)}]{Starobinsky:1979ty}
\bibinfo{author}{\bibfnamefont{A.~A.} \bibnamefont{Starobinsky}},
  \bibinfo{journal}{JETP Lett.} \textbf{\bibinfo{volume}{30}},
  \bibinfo{pages}{682} (\bibinfo{year}{1979}).

\bibitem[{\citenamefont{Caprini et~al.}(2010)\citenamefont{Caprini, Durrer, and
  Siemens}}]{Caprini:2010xv}
\bibinfo{author}{\bibfnamefont{C.}~\bibnamefont{Caprini}},
  \bibinfo{author}{\bibfnamefont{R.}~\bibnamefont{Durrer}}, \bibnamefont{and}
  \bibinfo{author}{\bibfnamefont{X.}~\bibnamefont{Siemens}},
  \bibinfo{journal}{\prd} \textbf{\bibinfo{volume}{82}},
  \bibinfo{pages}{063511} (\bibinfo{year}{2010}), \eprint{1007.1218}.

\bibitem[{\citenamefont{Nishizawa et~al.}(2009)\citenamefont{Nishizawa, Taruya,
  Hayama, Kawamura, and Sakagami}}]{nishizawa}
\bibinfo{author}{\bibfnamefont{A.}~\bibnamefont{Nishizawa}},
  \bibinfo{author}{\bibfnamefont{A.}~\bibnamefont{Taruya}},
  \bibinfo{author}{\bibfnamefont{K.}~\bibnamefont{Hayama}},
  \bibinfo{author}{\bibfnamefont{S.}~\bibnamefont{Kawamura}}, \bibnamefont{and}
  \bibinfo{author}{\bibfnamefont{M.-a.} \bibnamefont{Sakagami}},
  \bibinfo{journal}{Phys. Rev. D} \textbf{\bibinfo{volume}{79}},
  \bibinfo{pages}{082002} (\bibinfo{year}{2009}),
  \urlprefix\url{http://link.aps.org/doi/10.1103/PhysRevD.79.082002}.

\bibitem[{\citenamefont{Lee et~al.}(2008)\citenamefont{Lee, Jenet, and
  Price}}]{KJ}
\bibinfo{author}{\bibfnamefont{K.~J.} \bibnamefont{Lee}},
  \bibinfo{author}{\bibfnamefont{F.~A.} \bibnamefont{Jenet}}, \bibnamefont{and}
  \bibinfo{author}{\bibfnamefont{R.~H.} \bibnamefont{Price}},
  \bibinfo{journal}{\apj} \textbf{\bibinfo{volume}{685}}, \bibinfo{pages}{1304}
  (\bibinfo{year}{2008}),
  \urlprefix\url{http://stacks.iop.org/0004-637X/685/i=2/a=1304}.

\bibitem[{\citenamefont{Alves and Tinto}(2011)}]{alvestinto}
\bibinfo{author}{\bibfnamefont{M.~E. d.~S.} \bibnamefont{Alves}}
  \bibnamefont{and} \bibinfo{author}{\bibfnamefont{M.}~\bibnamefont{Tinto}},
  \bibinfo{journal}{Phys. Rev. D} \textbf{\bibinfo{volume}{83}},
  \bibinfo{pages}{123529} (\bibinfo{year}{2011}),
  \urlprefix\url{http://link.aps.org/doi/10.1103/PhysRevD.83.123529}.

\bibitem[{\citenamefont{{Detweiler}}(1979)}]{detweiler}
\bibinfo{author}{\bibfnamefont{S.}~\bibnamefont{{Detweiler}}},
  \bibinfo{journal}{\apj} \textbf{\bibinfo{volume}{234}}, \bibinfo{pages}{1100}
  (\bibinfo{year}{1979}).

\bibitem[{\citenamefont{Anholm et~al.}(2009)\citenamefont{Anholm, Ballmer,
  Creighton, Price, and Siemens}}]{anholm}
\bibinfo{author}{\bibfnamefont{M.}~\bibnamefont{Anholm}},
  \bibinfo{author}{\bibfnamefont{S.}~\bibnamefont{Ballmer}},
  \bibinfo{author}{\bibfnamefont{J.~D.~E.} \bibnamefont{Creighton}},
  \bibinfo{author}{\bibfnamefont{L.~R.} \bibnamefont{Price}}, \bibnamefont{and}
  \bibinfo{author}{\bibfnamefont{X.}~\bibnamefont{Siemens}},
  \bibinfo{journal}{Phys. Rev. D} \textbf{\bibinfo{volume}{79}},
  \bibinfo{pages}{084030} (\bibinfo{year}{2009}),
  \urlprefix\url{http://link.aps.org/doi/10.1103/PhysRevD.79.084030}.

\bibitem[{\citenamefont{Allen and Romano}(1999)}]{allenromano}
\bibinfo{author}{\bibfnamefont{B.}~\bibnamefont{Allen}} \bibnamefont{and}
  \bibinfo{author}{\bibfnamefont{J.~D.} \bibnamefont{Romano}},
  \bibinfo{journal}{Phys. Rev. D} \textbf{\bibinfo{volume}{59}},
  \bibinfo{pages}{102001} (\bibinfo{year}{1999}),
  \urlprefix\url{http://link.aps.org/doi/10.1103/PhysRevD.59.102001}.

\bibitem[{\citenamefont{Eardley et~al.}(1973)\citenamefont{Eardley, Lee,
  Lightman, Wagoner, and Will}}]{PRL30}
\bibinfo{author}{\bibfnamefont{D.~M.} \bibnamefont{Eardley}},
  \bibinfo{author}{\bibfnamefont{D.~L.} \bibnamefont{Lee}},
  \bibinfo{author}{\bibfnamefont{A.~P.} \bibnamefont{Lightman}},
  \bibinfo{author}{\bibfnamefont{R.~V.} \bibnamefont{Wagoner}},
  \bibnamefont{and} \bibinfo{author}{\bibfnamefont{C.~M.} \bibnamefont{Will}},
  \bibinfo{journal}{\prl} \textbf{\bibinfo{volume}{30}}, \bibinfo{pages}{884}
  (\bibinfo{year}{1973}),
  \urlprefix\url{http://link.aps.org/doi/10.1103/PhysRevLett.30.884}.

\bibitem[{\citenamefont{Tinto and Alves}(2010)}]{tintoalves2010}
\bibinfo{author}{\bibfnamefont{M.}~\bibnamefont{Tinto}} \bibnamefont{and}
  \bibinfo{author}{\bibfnamefont{M.~E. d.~S.} \bibnamefont{Alves}},
  \bibinfo{journal}{Phys. Rev. D} \textbf{\bibinfo{volume}{82}},
  \bibinfo{pages}{122003} (\bibinfo{year}{2010}),
  \urlprefix\url{http://link.aps.org/doi/10.1103/PhysRevD.82.122003}.

\bibitem[{\citenamefont{{Hellings} and {Downs}}(1983)}]{hellingsdowns}
\bibinfo{author}{\bibfnamefont{R.~W.} \bibnamefont{{Hellings}}}
  \bibnamefont{and} \bibinfo{author}{\bibfnamefont{G.~S.}
  \bibnamefont{{Downs}}}, \bibinfo{journal}{\apj}
  \textbf{\bibinfo{volume}{265}}, \bibinfo{pages}{L39} (\bibinfo{year}{1983}).

\bibitem[{\citenamefont{Manchester et~al.}(2005)\citenamefont{Manchester,
  Hobbs, Teoh, and Hobbs}}]{manchester-cat}
\bibinfo{author}{\bibfnamefont{R.~N.} \bibnamefont{Manchester}},
  \bibinfo{author}{\bibfnamefont{G.~B.} \bibnamefont{Hobbs}},
  \bibinfo{author}{\bibfnamefont{A.}~\bibnamefont{Teoh}}, \bibnamefont{and}
  \bibinfo{author}{\bibfnamefont{M.}~\bibnamefont{Hobbs}},
  \bibinfo{journal}{\apj} \textbf{\bibinfo{volume}{129}}, \bibinfo{pages}{1993}
  (\bibinfo{year}{2005}),
  \urlprefix\url{http://stacks.iop.org/1538-3881/129/i=4/a=1993}.

\bibitem[{\citenamefont{{Cordes} and {Lazio}}(2002)}]{cordes-lazio}
\bibinfo{author}{\bibfnamefont{J.~M.} \bibnamefont{{Cordes}}} \bibnamefont{and}
  \bibinfo{author}{\bibfnamefont{T.~J.~W.} \bibnamefont{{Lazio}}},
  \bibinfo{journal}{ArXiv Astrophysics e-prints}  (\bibinfo{year}{2002}),
  \eprint{arXiv:astro-ph/0207156}.

\end{thebibliography}

 \end{document}